%% file: paper.tex
\documentclass[11pt,a4paper]{article}
\usepackage{jheppub}
\usepackage{subfig}
\usepackage{soul}
\usepackage{amsfonts} 

\usepackage[normalem]{ulem}
\usepackage{xcolor}

\newcommand{\GeV}{~\mathrm{GeV}}
\newcommand{\TeV}{~\mathrm{TeV}}
\newcommand{\prevc}[1]{}

\newcommand{\winobino}[1]{}

\title{Effective field theory analysis of composite higgsino-like and wino-like thermal relic dark matter}
\author[a]{Ben Geytenbeek}
\author[a]{and Ben Gripaios}
\affiliation[a]{Cavendish Laboratory, J.J. Thomson Avenue, Cambridge, CB3 0HE, United Kingdom}
\emailAdd{bg364@cam.ac.uk}
\emailAdd{gripaios@hep.phy.cam.ac.uk}

\abstract{
We study the effective field theory (including operators up to
  dimension five) of models in which dark matter is composite, consisting of
  either an
  electroweak doublet Dirac fermion (`higgsino-like dark matter') or an
  electroweak triplet Majorana fermion (`wino-like dark
matter'). Some of the dimension-five operators in the former case cause mass splittings
between the neutralino and chargino states, leading to a depleted rate
of coannihilations and viable thermal relic dark matter with masses of
the order of tens to hundreds of$\GeV$ rather than the usual pure higgsino thermal relic
mass of $1\TeV$. No such
effects are found in the latter case (where the usual thermal relic
mass is $3\TeV$). Other operators, present for both wino- and higgsino-like dark matter, correspond to inelastic electromagnetic
dipole moment interactions and annihilation through these can lead to
viable models with dark matter masses up by an order of magnitude compared to the usual values.
}

\arxivnumber{}

\begin{document}
	
	\maketitle
	
	\input{introductionNOCOMMENTS}

	\input{models}
	\input{datasources}
	\input{results}
\subsection*{Acknowledgments}
We thank other members of the Cambridge Pheno Working Group for
discussions. Ben Gripaios is supported by STFC consolidated grant ST/P000681/1
and King’s College, Cambridge. Ben Geytenbeek is supported by the
Gates Cambridge Trust.
\IfFileExists{JHEP.bst}{
	\bibliographystyle{JHEP}
}{
	\bibliographystyle{references/JHEP}
}

\IfFileExists{references.bib}{
	\bibliography{references}
}{
	\bibliography{references/references}
}
	
\end{document}

%% file: introductionNOCOMMENTS.tex
\section{Introduction}

	The existence and identity of the dark matter in the universe have long been major issues in astronomy and particle physics~\cite{zwicky33}. There is now an overwhelming catalogue of astronomical evidence for the existence of dark matter, including from galactic rotation curves~\cite{roberts75}, large scale galaxy structure~\cite{white78}, the cosmic microwave background~\cite{spergel03}, and colliding galaxy clusters~\cite{clowe04}, but its identity remains a mystery.

	A common and popular proposal to explain the particle nature of dark matter (assuming it is indeed a particle) is a fermion (Majorana or Dirac), which must of course be electrically neutral and an $SU(3)$ color singlet, but may have some electroweak charge. The simplest possibilities are a Majorana singlet or triplet, or a doublet (which must be Dirac). Though we may generalise to larger multiplets~\cite{cirelli06}, these examples have the nice feature that they correspond to the bino, wino, and higgsino of supersymmetric models. Even though supersymmetry will play no role in this work, we will freely employ its familiar nomenclature, referring, for example, to the electrically-neutral states in a multiplet as neutralinos, and to the charged states as charginos.

	The models with a higgsino or a wino are particularly attractive because coannihilations~\cite{griest91,mizuta93,drees97}, namely annihilation processes between the neutralinos and charginos, and the Sommerfeld enhancement~\cite{hisano07} lead to natural thermal relic dark matter candidates close to the TeV scale. To wit, doublet models lead to a $1 \TeV$ thermal relic while triplet models lead to a $3\TeV$ thermal relic~\cite{cirelli06, arkani-hamed06,profumo04,baer11,baer13,bae15,bharucha17,roszkowski18}.

	Unfortunately, neither of these models are viable in their simplest renormalizable incarnations, because of the many existing searches for dark matter (see e.g. \cite{kahlhoefer17,lux17,xenon17,fermilat15,ams16}).  The pure higgsino is ruled out by direct detection due to the near degeneracy with the charged state (see, {\em e.g.}~\cite{badziak17}) while winos are largely ruled out by indirect detection due to their strong tree-level annihilation into a pair of $W$ bosons~\cite{fan13,cohen13}.

	At least in the case of the higgsino, there is an easy way out: even a tiny splitting of at least around 100 keV in the masses relaxes the direct detection bounds as the scattering rates are now suppressed because the process is inelastic~\cite{smith01,tuckersmith05} (for recent discussions of bounds in the SUSY case, see~\cite{bramante16,baer18, kowalska18,krall18}).  New mass terms which split the neutralino and chargino masses also reduce the rate of coannihilations of dark matter, giving an increased relic density~\cite{griest91,mizuta93,drees97}.
	Such a tiny mass splitting can arise from new physics at much higher scales,\footnote{For example, in SUSY models, the splitting can arise due to radiative corrections from loops of heavy squarks~\cite{pierce94a,pierce94b,lahanas94,drees97} or integrating out higher scale neutralinos~\cite{drees97,howe12}. For other UV-complete models generating mass splittings, see \cite{profumo04,gherghetta99,feng99,cheng99,ibe13,yamada10}.} in which case it can be described in a completely model-independent fashion by dimension five operators in an effective field theory description~\cite{ellis00,hisano05,nagata15,bharucha17}. In other words, we regard dark matter as being composite.\footnote{For recent related work on composite dark matter, see \cite{Redi:2020qyf}.}

	We thus see that there is a simple, viable description of dark matter, in terms of a Lagrangian containing only the Standard Model plus a higgsino, with terms of dimension up to and including five. Our goals in this work are, firstly, to explore the full parameter space of this model (with all possible dimension-five operators added), which we call higgsino-like dark matter and, secondly, to do the same for the corresponding model with the higgsino replaced by a wino, which we call wino-like dark matter.

	There are two general classes of operators in each model.\,One\,class\,features\,a\,dimension-5 coupling between the dark matter and the Higgs field. Such operators can produce mass splittings between the neutral and charged components after the Higgs field is spontaneously broken~\cite{nagata15,hisano15,bharucha17}.
	The other class of operators correspond to electric and magnetic dipole moments for dark matter~\cite{nagata15b}. Such operators are have been well-studied at the GeV scale~\cite{pospelov00,barger11,sigurdson04,masso09,fitzpatrick10,fortin12,heo10,heo11,delnobile12,delnobile14,gresham14,mohanty15,nagata15b,geytenbeek17} but here we analyse their effects near the TeV scale.

	In section~\ref{sec:models}, we will introduce the formalism for each of the relevant dimension-5 operators. With these operators added, in section~\ref{sec:results}, we investigate the effect on the relic density and also compare to direct detection, indirect detection and collider bounds to test the viability of each of the relevant regions, while taking care to ensure that we remain within the region of validity of the effective field theory description. We show that higgsino-like models with a sizeable mass splitting can produce viable thermal relics down to a few hundred GeV, whilst direct annihilation through these operators can produce viable thermal relics at scales of up to tens of TeV. For wino-like models, the only viable thermal relics are at scales of up to tens of TeV.

%% file: models.tex
\section{Models}
\label{sec:models}

	\subsection{Doublet models (higgsinos)}

	Consider introducing a pair of Majorana spin-1/2 $SU(2)$ doublets to the Standard Model. Though we borrow the notation and terminology of higgsinos from supersymmetric models, we seek to retain generality and so do not seek to specify the higher-scale model.	
	The particles are represented by a pair of doublets of two-component spinors $\widetilde{H}_u = \begin{pmatrix}\widetilde{H}_u^+ \\\widetilde{H}_u^0\end{pmatrix}$ and $\widetilde{H}_d = \begin{pmatrix}\widetilde{H}_d^0 \\\widetilde{H}_d^-\end{pmatrix}$, where we have used the round brackets for the representation of the gauge group.

	 The mass term in the Lagrangian is given by
	\begin{equation}
		\mathcal{L} \supset -\mu \epsilon^{ij}(\widetilde{H}_u)_i(\widetilde{H}_d)_j + \rm{h.c.}\, ,
	\end{equation}
	for the antisymmetric tensor $\epsilon^{ij}$ such that $\epsilon^{12} = - \epsilon^{21} = +1$, where the indices $ij$ to indicate the gauge components. Note here that we can construct a pair of Dirac fermions with four-component spinors in the chiral basis $\widetilde{H}^0 = \begin{bmatrix} \widetilde{H}_u^0 \\ \widetilde{H}_d^{0\dagger}\end{bmatrix}$ and $\widetilde{H}^+ = \begin{bmatrix} \widetilde{H}_u^+\\ \widetilde{H}_d^{-\dagger}\end{bmatrix}$ where the square brackets indicate spinor components. Hence, the mass term may also be written as
	\begin{equation}
		\mathcal{L} \supset -\mu \overline{\widetilde{H}^0} \widetilde{H}^0 + \mu \overline{\widetilde{H}^+}\widetilde{H}^+.
	\end{equation}
	We can factor the Dirac spinors into a single $SU(2)$ doublet $\widetilde{H} = \begin{pmatrix} \widetilde{H}^+ \\ \widetilde{H}^0 \end{pmatrix}$, The model therefore also represents a single $SU(2)$ doublet being added to the Standard Model. The mass is given by
	\begin{equation}
		\label{eq:higgsdiracmass}
		\mathcal{L} \supset -\mu \overline{\widetilde{H}_i}\widetilde{H}_i.
	\end{equation}

	The kinetic term for the Dirac doublet appears in the Lagrangian interacting under the $SU(2)\times U(1)$ gauge symmetry of the Standard Model
	\begin{equation}
		\label{eq:higgsinodim4}
		\mathcal{L} \supset i\overline{\widetilde{H}_i}\gamma^{\mu}\left(\partial_\mu-igA^a_{\mu}\frac{\tau_{ij}^a}{2}-ig'YB_{\mu}\delta_{ij}\right)\widetilde{H}_j
		,
	\end{equation}
	where $A^a_{\mu}$ and $B_{\mu}$ are, respectively, the $SU(2)$ and $U(1)$ Standard Model gauge bosons in the interaction basis, $g$ and $g'$ are the Standard Model gauge couplings, $Y = +\frac{1}{2}$ is the weak hypercharge of the doublet, and $\tau_{ij}^a$ are the generators of the $SU(2)$ symmetry in the fundamental representation. Under the spontaneous symmetry breaking of the Higgs, we get the standard expressions of the mass-basis gauge bosons in the usual way:
	\begin{align}
		W^{\pm}_\mu \equiv&~ \frac{1}{\sqrt{2}} (A_\mu^1 \mp i A_\mu^2),
		\\
		Z_\mu \equiv&~ \frac{-gB_\mu + gA_\mu^3}{\sqrt{g^2+g'^2}},
		\\
		A_\mu^\gamma \equiv&~\frac{gB_\mu+g'A_\mu^3}{\sqrt{g^2+g'^2}}.
	\end{align}
	We have now recovered the usual interactions between an $SU(2)$ fermionic doublet and the gauge bosons. The mass and covariant derivative terms are the only interactions allowed by the new doublets; there are no new gauge-invariant Yukawa terms in the theory as the only scalar particle in the model is the Higgs boson. 
	
	The only free parameter in this model is $\mu$, the mass parameter of the higgsinos. By imposing that the model must be a thermal relic dark matter candidate, then $\mu$ is constrained to a mass of $1 \TeV$~\cite{baer16,baer11,baer13,fan13,ellis98,profumo04,bramante16}.

In our description, any additional physics above the scale of $\mu$
is encoded in higher-dimensional effective field theory operators, which can result in modifications to both the masses and couplings of the higgsinos. Examples of such physics include the top-stop-higgsino Yukawa from the MSSM, whose effects are dependent on the mass of the stop squark, or the addition of more weakly-interacting neutral fermions at higher mass scales (such as the bino or sneutrinos in the MSSM). The key advantage of following an effective field theory approach is that we do not need to specify the exact nature of the higher scale physics. In particular, supersymmetry is a possible, but not the only, UV completion of the theory.

	In this work, we consider the eight independent dimension-5 operators for the pair of (Majorana) $SU(2)$ doublet higgsinos. Four such operators correspond to an interaction with a scalar Higgs Boson $H = \begin{pmatrix} H^+ \\ H^0 \end{pmatrix}$ as discussed in ref.~\cite{nagata15,hisano15,bharucha17}, namely
	\begin{align}
		\label{eq:OH1}
		\mathcal{O}_1^H = &~ (H^\dagger)^i (\widetilde{H}_u)_i (H^\dagger)^j (\widetilde{H}_u)_j \, ,
		\\
		\label{eq:OH2}
		\mathcal{O}_2^H = &~ \epsilon^{ij}\epsilon^{kl}(H)_i(\widetilde{H}_d)_j(H)_k(\widetilde{H}_d)_l \, ,
		\\
		\label{eq:OH3}
		\mathcal{O}_3^H = &~ \epsilon^{jk}(H^\dagger)^{i}(\widetilde{H}_u)_i(H)_j(\widetilde{H}_d)_k \, ,~\text{and}
		\\
		\label{eq:OH4}
		\mathcal{O}_4^H = &~
		\epsilon^{jk}(H^\dagger)^{i}(\widetilde{H}_d)_i(H)_j(\widetilde{H}_u)_k \, .
	\end{align}
	As noted in ref.~\cite{nagata15}, no terms of the form $\epsilon^{ij}\epsilon^{kl}(H)_i(H)_j (\widetilde{H}_d)_k (\widetilde{H}_d)_l$ are permitted due to the fact that the $H$ bosons are symmetric. Given the two-dimensional $SU(2)$ generators $\tau^{a}_{ij}$, it is possible to write these operators in the form $(H_i^\dagger \tau^a_{ij} \widetilde{H}_j)(H_{k}^\dagger \tau^a_{kl} \widetilde{H}_l)$ using the Fierz relation of the generators $\tau_{ij}^a \tau_{kl}^a = 2\delta_{il}\delta_{jk} - \delta_{ij}\delta_{kl}$. 
	
	Terms like these arise from, for example, integrating the gauginos out from the MSSM. They naturally allow for modifications to the higgsino masses after the Higgs boson obtains a vacuum expectation value. Physically, the mass modification represents a small mixing of the masses to a significantly heavier state.
	
	In addition, we introduce four operators which together contain the electric and magnetic dipole moments of the higgsinos, namely
	\begin{align}
		\label{eq:OH5}
		\mathcal{O}_5^H = &~ \overline{\widetilde{H}_i} \sigma^{\mu\nu}B_{\mu\nu} \widetilde{H}_i \, ,
		\\
		\label{eq:OH6}
		\mathcal{O}_6^H = &~ \overline{\widetilde{H}_i} \tau^a_{ij}\sigma^{\mu\nu}W_{\mu\nu}^a \widetilde{H}_j \, , 
		\\
		\label{eq:OH7}
		\mathcal{O}_7^H = &~\overline{\widetilde{H}_i} \sigma^{\mu\nu}B_{\mu\nu}^* \widetilde{H}_i \, ,~\text{and}
		\\
		\label{eq:OH8}
		\mathcal{O}_8^H = &~\overline{\widetilde{H}_i}
		\tau_{ij}^a \sigma^{\mu\nu}(\widetilde{W}^a)^*_{\mu\nu} \widetilde{H}_j \, ,
	\end{align}
	 where $B_{\mu\nu}$ and $W^a_{\mu\nu}$ are the field strength tensors for the $U(1)$ and $SU(2)$ gauge fields respectively
	\begin{align}
		B_{\mu\nu} = &~ \partial_\mu B_\nu - \partial_\nu B_\mu \,,
		\\
		W^a_{\mu\nu} = &~ \partial_\mu A_\nu^a - \partial_\nu A_\mu^a + f^{abc} A_\mu^b A_\mu^c \, ,
	\end{align}
	and $B_{\mu\nu}^*$ and $(W^a)_{\mu\nu}^*$ are the dual field strength tensors $B_{\mu\nu}^* = \epsilon_{\mu\nu\sigma\rho} B^{\sigma\rho}$ and $(W^a)_{\mu\nu}^* = \epsilon_{\mu\nu\sigma\rho} (W^a)^{\sigma\rho}$. Note here that the $\tau^a_{ij}$  term in eqs.~(\ref{eq:OH6}) and (\ref{eq:OH8}) mixes the states of the doublet. As the neutral states are mixed, then if there is a mass splitting the resultant dipole moments will be inelastic~\cite{smith01}.

The collection of operators in eqs.~(\ref{eq:OH1}-\ref{eq:OH8}) represent the complete set of gauge-invariant operators at dimension 5 up to the usual redundancies in effective field theory, as dimensional analysis and symmetry restrict us to operators involving two fermions and two bosons.

	The neutral components of operators $\mathcal{O}_5^H$ and $\mathcal{O}_6^H$ correspond to an electroweak magnetic dipole operator, and the neutral components of operators $\mathcal{O}_7^H$ and $\mathcal{O}_8^H$ correspond to an electroweak electric dipole operator, as can be seen using the identity $\frac{i}{2}\sigma^{\sigma\rho}\epsilon_{\mu\nu\rho\sigma} = \sigma^{\mu\nu}\gamma^5$. 	\footnote{It is well known that Majorana particles cannot possess electric or magnetic dipole moments; the first permissible electromagnetic term is the dimension-6 anapole moment~\cite{ho13}. The electric and magnetic dipoles are both $CPT$-odd, so for a particle that is $CPT$ self-conjugate, the terms cannot exist in a $CPT$ consistent theory~\cite{radescu85}. However, the restriction does not apply in this theory because of the presence of two $SU(2)$ doublets being introduced simultaneously. Where both of the doublets have identical masses, the accounting trick of making the 2 doublets into one Dirac fermion in eq.~\ref{eq:higgsdiracmass} resolves the issue. Where the masses are split via operators $\mathcal{O}_1^H$ and $\mathcal{O}_2^H$, these equations describe an inelastic dipole between the two mass eigenstates, or between the neutralino eigenstates and the chargino eigenstates. } Electric and magnetic dipolar dark matter have been studied extensively in the literature for electromagnetically interacting $U(1)$ models at masses of a few to a hundred GeV ~\cite{pospelov00,barger11,sigurdson04,masso09,fitzpatrick10,fortin12,heo10,heo11,delnobile12,delnobile14,gresham14,mohanty15,geytenbeek17}. 
	Our treatment here considers an electroweak $SU(2)\times U(1)$ model, which also allows interactions between the charged higgsinos and $W^\pm$ bosons. Also, note that the electric dipole-like terms with the dual field strength tensors are $CP$-violating terms.

	 To examine the properties of these operators, we collate the dimension-5 terms into the effective Lagrangian
	\begin{equation}	
		\mathcal{L} \supset \sum_{i=1}^8 \frac{c_i}{\Lambda} \mathcal{O}_i^H \, + \mathrm{h.c.} \,,
	\end{equation}
	for some coupling coefficients $c_i$ and UV-cutoff scale $\Lambda$. In principle, some of the $c_i$ may be complex, but here we will assume them to be all real.
	
	As eqs.~(\ref{eq:OH1}-\ref{eq:OH4}) give four operators that provide a coupling to the Higgs boson, these operators will provide additional mass terms to $\widetilde{H}$ after electroweak symmetry breaking~\cite{nagata15}.
	Notably, $\mathcal{O}_1^H$ and $\mathcal{O}_2^H$ create a splitting of the masses of the two neutral states by placing terms on the main diagonal of an otherwise anti-diagonal mass matrix. With these terms switched on, the magnitude of the eigenvalues of the mass matrix are no longer equal. $\mathcal{O}_3^H$ adds off-diagonal terms, because only the neutral component of the Higgs gets a vev. Finally, $\mathcal{O}_4^H$ shifts the mass of the charged state. Hence we obtain the mixing matrix
	\begin{equation}
		\mathcal{L} \supset \widetilde{H}^0_i\mathcal{M}_{ij} \widetilde{H}^0_j = -\frac{1}{2} \begin{pmatrix} \widetilde{H}_u & \widetilde{H}_d \end{pmatrix}
		\begin{pmatrix} \frac{v^2 c_1}{\Lambda} & -\mu+\frac{v^2 c_3}{2\Lambda} \\ -\mu + \frac{v^2 c_3}{2\Lambda} & \frac{v^2 c_2}{\Lambda} \end{pmatrix}
		\begin{pmatrix} \widetilde{H}_u \\ \widetilde{H}_d \end{pmatrix} \, .
	\end{equation}
	The mass matrix is diagonalised for complex parameters $\mu$, $c_1$, $c_2$ and $c_3$ through the usual diagonalization of complex symmetric matrices~\cite{choi07}; the masses obtained are given by
	\begin{equation}
	\label{eq:msquared}
		m_{1,2}^2 = \frac{1}{2}\left( 2 |\bar{\mu}|^2 + \frac{v^4(|c_1|^2+|c_2|^2)}{\Lambda^2} \pm  \sqrt{\frac{v^8(|c_1|^2-|c_2|^2)^2}{\Lambda^4} + 4\frac{v^4|c_2\bar{\mu}^*+c_1^*\bar{\mu}|^2}{\Lambda^2}}\right) \, .
	\end{equation}
	In the limit that the parameters are real (more specifically, that the complex phases are equal), the neutral masses are given by 
	\begin{align}
	\label{eq:m1}
		m_1 = &~\left|\widetilde{\mu} - \frac{c_1+c_2}{2\Lambda}v^2 \right|\, 
	\intertext{and}
	\label{eq:m2}
		m_2 = &~\left|\widetilde{\mu} + \frac{c_1+c_2}{2\Lambda}v^2 \right|\, ,
	\intertext{with the charged mass given by}
	\label{eq:mpm}
		m_\pm =&~ \left|\mu + \frac{c_4}{2\Lambda}v^2\right| \, .
	\end{align}
	Here we have defined
	\begin{align}
		\label{eq:mutilde}
		\widetilde{\mu} =&~ \frac{1}{2}\sqrt{\frac{(c_1-c_2)^2}{\Lambda^2}v^4 + 4\bar{\mu}^2} \, ,
	\intertext{and}
		\label{eq:mubar}
		\bar{\mu} =&~ \mu- \frac{c_3}{2\Lambda}v^2 \, ,
	\end{align}
	where $v \simeq 246\GeV$ is the Higgs vacuum expectation value.  
	Note that in the limit where $c_1\simeq c_2$ as we consider later in this work, then $\widetilde{\mu}\simeq \bar{\mu}$. Finally, the transformation from the interaction basis to the mass basis is given by
	\begin{equation}
		\begin{pmatrix}
			\widetilde{\chi}_1^0 \\
			\widetilde{\chi}_2^0
		\end{pmatrix}
		=
		\begin{pmatrix}
			\cos\theta & -\sin\theta \\
			\sin\theta & \cos\theta
		\end{pmatrix}
		\begin{pmatrix}
			\widetilde{H}_d^0 \\
			\widetilde{H}_u^0
		\end{pmatrix} \, ,
	\end{equation}
	with
	\begin{equation}
		\tan\theta \simeq 1 + \frac{c_2-c_1}{2\mu\Lambda}v^2.
	\end{equation}

	It must be noted here that eqs.~(\ref{eq:m1}) and (\ref{eq:m2}) imply that, in the case where $\widetilde{\mu} \sim \frac{v^2}{\Lambda}$, the mass of the lighter particle approaches zero, while the heavier of the two neutralinos and the chargino remain of $\mathcal{O}(\mu)$. Thus a large mass difference between the two masses can be generated, which is crucial for the effect on the relic density. Where there is a sizeable mass difference between the chargino and neutralino, the rate of coannihilations is significantly reduced, greatly impacting the thermal relic density. This region is capable of producing feasible thermal relic dark matter with significantly altered dark matter mass.

	\subsection{Triplet models (winos)}

	Next, we introduce a weakly-interacting, Majorana, $SU(2)$-triplet, $\widetilde{W}$, with zero weak hypercharge. The model is analogous to a supersymmetric wino, and again we borrow the notation and terminology, but stress that such a model need not be supersymmetric. The components are represented in the usual way for winos, with two charged components $\widetilde{W}^\pm$ and a neutral component $\widetilde{W}^0$, which may be written in doublet form as $\widetilde{W}_i = \left(\widetilde{W}^+,\widetilde{W}^0,\widetilde{W}^-\right)$. The mass term is given by
	\begin{equation}
		\mathcal{L} \supset -\frac{1}{2} M_2 \widetilde{W}_i \widetilde{W}_i \, .
	\end{equation}
	Since there is zero hypercharge, there is only an interaction term corresponding to the $SU(2)$ symmetry of the Standard Model
	\begin{equation}
		\mathcal{L} \supset i \widetilde{W}_i \overline{\sigma}^\mu \left(\delta_{ij}\partial_\mu - igA_\mu^a \frac{T_{ij}^a}{2}\right)\widetilde{W}_j \, +\mathrm{h.c.},
	\end{equation}
	where $T_{ij}^a$ are the three-dimensional representations of the generators of $SU(2)$ and $\sigma^\mu$ is the . These dimension-4 interactions can produce a thermal relic particle with mass $3 \TeV$~\cite{arkani-hamed06,profumo04,baer11,baer13,bae15,fan13}. Variations to the mass of the relic occur if the neutral component of the triplet is mixed with a singlet or doublet (i.e. a bino-like or higgsino-like particle). As for the higgsino, we seek to model the effect of any such (heavy)  additional new physics through effective field theory operators. 
	
	Dimensional analysis and symmetry restrict us at dimension 5 to a coupling between two fermionic states and two bosonic states. As for the doublet case, we expect a coupling to the Higgs bosons of the form
	\begin{equation}
		\label{eq:winohiggsnosplitting}
		\mathcal{O}_1^W = (H^\dagger)^i(H)_i \widetilde{W}^2 \, .
	\end{equation}
	The operator in eq.~(\ref{eq:winohiggsnosplitting}) contains two Higgs bosons, and hence we expect to see shifts or splittings in the mass spectrum after electroweak symmetry breaking. Unlike the doublets, where the masses occur on the anti-diagonal of the mixing matrix, here we have a direct contribution to the doublet masses
	\begin{equation}
		\mathcal{L} \supset \widetilde{W}^i \left(M_2 + \frac{d_1 v^2}{\Lambda}\right) \widetilde{W}^i \, .
	\end{equation}
	Here, the chargino and neutralino masses are equal regardless of the scale of the dimension-5 coupling.

	For the interaction between the triplet fermion and the gauge bosons at dimension 5, it is possible to write down two  gauge-invariant terms, connecting the $\widetilde{W}$ fields either to the Abelian $B_{\mu\nu}$ field strength tensor, and the non-Abelian $W_{\mu\nu}$ field strength tensor. 
Each of these terms also has an equivalent term with the dual field strength tensor which is $CP$ violating. 

The two terms are
\begin{align}
\label{eq:OW3}
\mathcal{O}_2^W =&~ \widetilde{W}_i T_{ij}^a \sigma^{\mu\nu} W_{\mu\nu}^a \widetilde{W}_j \, ,~\text{and}
\\
\label{eq:OW4}
\mathcal{O}_3^W =&~ \widetilde{W}_i T_{ij}^a \sigma^{\mu\nu} (W^a_{\mu\nu})^* \widetilde{W}_j \, .
\end{align}

We cannot introduce operators containing $\overline{W}_i\sigma^{\mu\nu}B_{\mu\nu}\overline{W}_i$ or $\overline{W}_i\sigma^{\mu\nu}B_{\mu\nu}^*\overline{W}_i$ as though they are Lorentz invariant, they vanish identically via the antisymmetry of the fermionic components. Indeed, these components would have corresponded to an electric dipole operator which is forbidden for Majorana particles under a $CPT$-invariant theory~\cite{ho13,radescu85}. 

However, by introducing a gauge field with matrix components, it is possible to form a Lorentz invariant via a coupling between the antisymmetric components of $T_{ij}^a$ and the antisymmetric combination $\overline{W}_i\overline{W}_j$. Hence the terms in eqs.~(\ref{eq:OH3}) and (\ref{eq:OH4}) can avoid the constraints on magnetic dipoles in $CPT$-invariant theories.

Finally, the Lagrangian is compiled as
\begin{equation}
\mathcal{L} \supset \sum_{i=1}^3 \frac{d_i}{\Lambda} \mathcal{O}_i^W + \mathrm{h.c.} \, ,
\end{equation}
for coupling parameters $d_i$ and UV-cutoff $\Lambda$. As usual in EFT, we take the coefficients $c_i$ to be order one or smaller. In principle some may be complex, but here we take them all to be real.

We have now introduced all dimension-5 terms of the higgsinos and winos introduced separately into the theory. These terms account quite generally for the leading effects of physics at higher scales. The new terms provide both electric and magnetic dipole-like interactions, and couplings to the Higgs provide mass terms. There is a regime where if the effective field theory couplings is sufficiently large, the masses may split, cancel or dominate, which is relevant to the coannihilation rates in the relic density calculations. 

%% file: datasources.tex
\section{Relic density}
\label{sec:results}
	
	Any reasonable dark matter model must account for the abundance of dark matter in the universe. In the thermal relic dark matter scenario, the dark matter is in thermal equlibrium with the Standard Model in the early universe. At the freeze-out epoch, the expansion of the universe prevents dark matter from annihilating, and the remaining relic dark matter density remains constant through to the present day. The thermal relic density is dependent on the size of the annihilation cross-section. Although there are models which explain the abundance of dark matter through non-thermal mechanisms~\cite{baer15}, in this work we focus solely on thermal models.
	
	In higgsino-like or wino-like dark matter, the only free parameter affecting the annihilation cross section at the renormalizable level is the mass parameter $\mu$ or $M_{2}$,  respectively. Here, the only relevant interactions that can contribute to the dark matter annihilation in the early universe are with the gauge bosons. In the Lagrangian, these interactions arise in the covariant derivative, and lead to, for example, $s$-channel annihilation via a $Z$ boson or $t$-channel annihilation to a pair of $W$ bosons. The coupling strength is therefore given by the electroweak coupling parameters $g$ and $g'$, which are fixed in the Standard Model. 
	
	As the annihilation cross section increases, the relic density decreases as the dark matter abundance is depleted more efficiently. Hence, aside from resonance effects near the masses of the gauge bosons, there is a direct proportionality between the relic density and the dark matter mass parameters. The relic density measurement then uniquely constrains the relevant mass parameter. For higgsino-like dark matter at dimension 4 and below, the relic density constrains the mass to around $1 \TeV$, whereas wino-like dark matter at dimension 4 and below is constrained to around $3 \TeV$. Searches for neutralino dark matter have focussed on these mass regions, with the $3 \TeV$ thermal relic wino dark matter being ruled out by indirect detection experiments~\cite{fan13,cohen13},  with the $1 \TeV$ thermal higgsino requiring mass splittings, albeit small ones, in order not to be ruled out by direct detection experiments.

	In this work, we investigate the effect of introducing the eight additional dimension-5 operators for higgsino-like dark matter in eqs.~(\ref{eq:OH1}-\ref{eq:OH8}) or the five dimension-5 operators for wino-like dark matter in eqs.~(\ref{eq:winohiggsnosplitting}), (\ref{eq:OW3}) and (\ref{eq:OW4}) to the relic density of the respective particles. In particular, we consider the effect on the relic density spectrum owing to the modifications to the higgsino-like masses in eqs.~(\ref{eq:m1})~and~(\ref{eq:m2})\winobino{ and the wino-like masses in eqs.~(\ref{eq:Wm1}) and (\ref{eq:Wm2})}. We calculate the relic densities using the \texttt{micrOMEGAs} computer code~\cite{belanger02,belanger06,barducci18}, using the inbuilt \texttt{CalcHEP}~\cite{belyaev13} and \texttt{LanHEP}~\cite{semenov08} functionality to generate the Feynman diagrams and associated squared matrix elements.  
	
	For each operator we consider, we vary two parameters that contribute to the mass spectrum of the higgsino-like or wino-like particle. These are the mass parameter $\mu$ or $M_{1,2}$ and the UV cut-off associated with the dimension-5 coupling $\Lambda$. In all cases, we set $c_i, d_i = \pm1$, which represents roughly the largest possible allowed coupling magnitude. For each of the operators, there is a region of the parameter space where the masses are greater than the cut-off scale. In these regions, the effective field theory is not valid, in that the details of the higher-scale physics become relevant; we exclude such regions from the results.
	
	For each of our models, we consider the relic density and constraints from experimental results. We use compare the simulated relic density to the observed relic density from WMAP and Planck of $\Omega h^2 = 0.1188\pm0.0010$~\cite{komatsu11,planck16}. The regions which provide the correct relic density are shown in each of the figures as a thick green line. A further constraint arises from collider experiments, notably the invisible width of $Z$ boson decays as measured by LEP~\cite{l389,aleph89,opal89,delphi89,baer90,drees88}. The mass of the charged component of the higgsino-like or wino-like fermions must be greater than $37\GeV$~\cite{baer90} to avoid a contribution to $\Gamma_Z$. However, no such constraint can be placed on the neutral component of the higgsino-like fermion, as there is no direct coupling to the $Z$ boson.

	There are also indirect detection constraints imposed by astronomical observations. Searches for dark matter using gamma ray astronomy at the TeV scale include satellite-based (e.g. Fermi-LAT) and ground-based (e.g. HESS) telescopes looking for the by-products of dark matter annihilation in both the galactic centre~\cite{ackermann12} and extra-galactic sources~\cite{fermilat14}.  In this work we impose the Fermi-LAT dwarf spheriod~\cite{fermilat14} and HESS galactic centre constraints~\cite{hess16}; the former bounds are stronger for masses of around $100\GeV$, the latter is stronger for masses above $1\TeV$. Before the introduction of dimension-5 operators, the Fermi-LAT constraints rule out wino-like thermal dark matter~\cite{fan13}. The strongest bound for the present work is in the $W^+W^-$ channel, as it is directly produced by the electric and magnetic dipole interactions in eqs.~(\ref{eq:OH5}-\ref{eq:OH8}) and eqs~(\ref{eq:winohiggsnosplitting}), (\ref{eq:OW3}) and (\ref{eq:OW4}). We calculate the annihilation rates for each point in our parameter space using the \texttt{micrOMEGAs} code~\cite{belanger02,belanger06,barducci18}, followed by a calculation of the Sommerfeld enhancement~\cite{slatyer10,feng10,cassel10}. All operators are affected by the constraints on the $W^+W^-$ annihilation channel when the mass of the LSP is near the resonance at the $W$ boson mass, regardless of the dimension-5 couplings.
		
	The direct detection bounds on the higgsino-like models are weak at tree level: the $t$-channel scattering is suppressed as the two neutral components are of near equal mass creating a cancellation in the vertex with the $Z$ boson
and there is no $s$-channel annihilation via a heavy squark as there is in supersymmetric models~\cite{hisano11,hill12}. The result is an expected scattering cross section below the neutrino floor for direct detection. The dimension-5 operators which do have a coupling to the $Z$ boson also undergo similar cancellations at tree level.	Meanwhile, for the wino-like models there are no diagrams at tree level that permit elastic scattering of the dark matter off nucleons.
	In general, direct detection bounds are not strong enough to rule out any significant regions of the parameter space of the higgsino-like particles, at least under the usual conditions where there is a tiny mass splitting between the higgsino and the chargino~\cite{smith01,tuckersmith05}. The exception occurs for regions where the LSP is the chargino, that is, when $\frac{c_1+c_2}{2\Lambda}v^2 > \mu+ \frac{c_4}{2\Lambda}v^2$. For the higgsino, this corresponds to small $\mu$ and a small cut-off scale. Such regions are strongly excluded as they represent charged dark matter, which has been excluded by numerous experiments, for example, the abdundance of superheavy isotopes of hydrogen in sea water~\cite{verkerk92}.
	
	\subsection{Higgsino electric dipole interactions}
	
	\begin{figure}[t]
		{
			\subfloat[$\frac{1}{\Lambda}\mathcal{O}_7^H$ from eq.~(\ref{eq:OH7})\label{fig:cBCP}]{\includegraphics[width=0.49\textwidth]{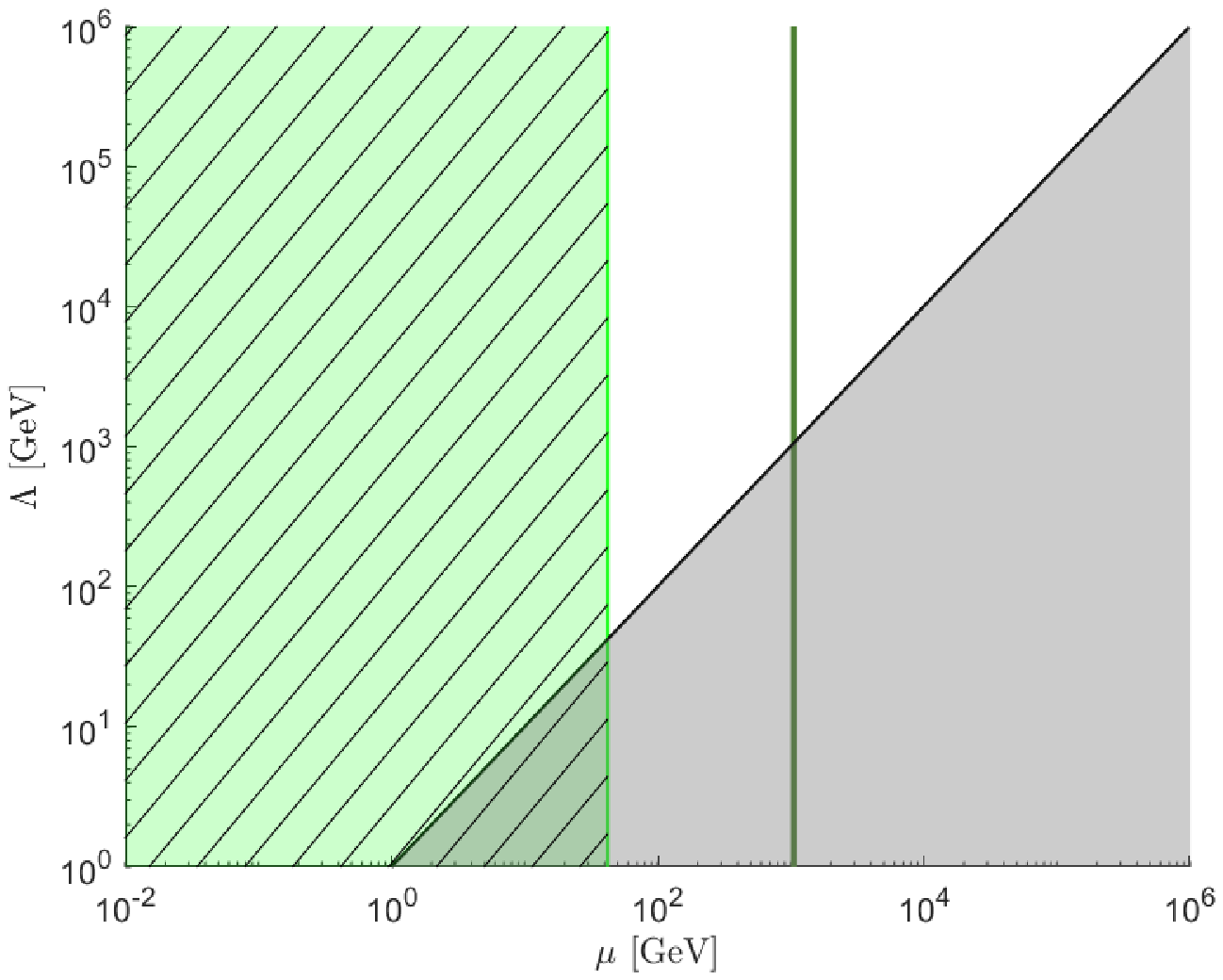}}
			\subfloat[$\frac{1}{\Lambda}\mathcal{O}_8^H$ from eq.~(\ref{eq:OH8})\label{fig:cWCP}]{\includegraphics[width=0.49\textwidth]{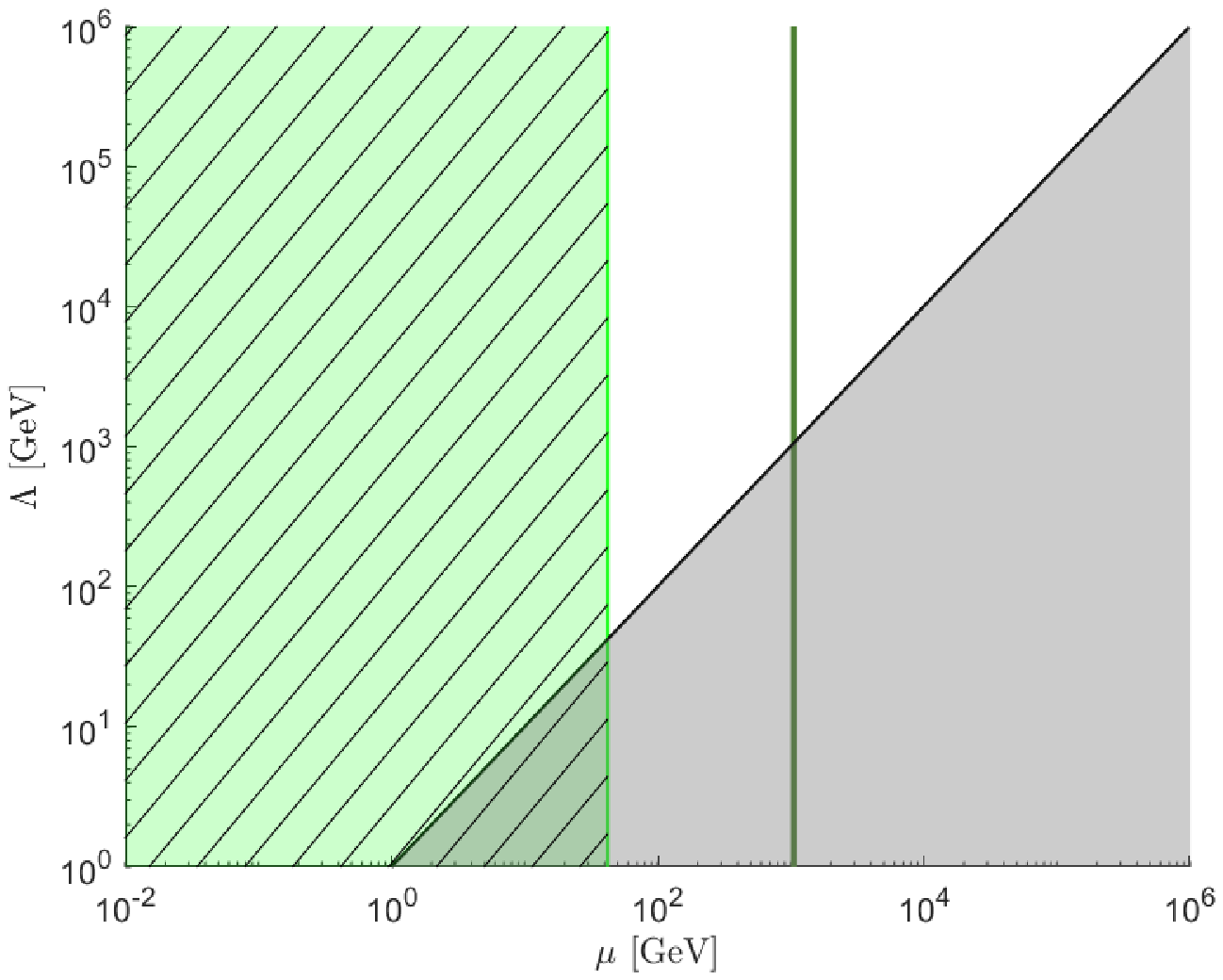}}
		}
		{
			\subfloat[$\frac{1}{\Lambda}\mathcal{O}_5^H$ from eq.~(\ref{eq:OH5})\label{fig:cB}]{\includegraphics[width=0.49\textwidth]{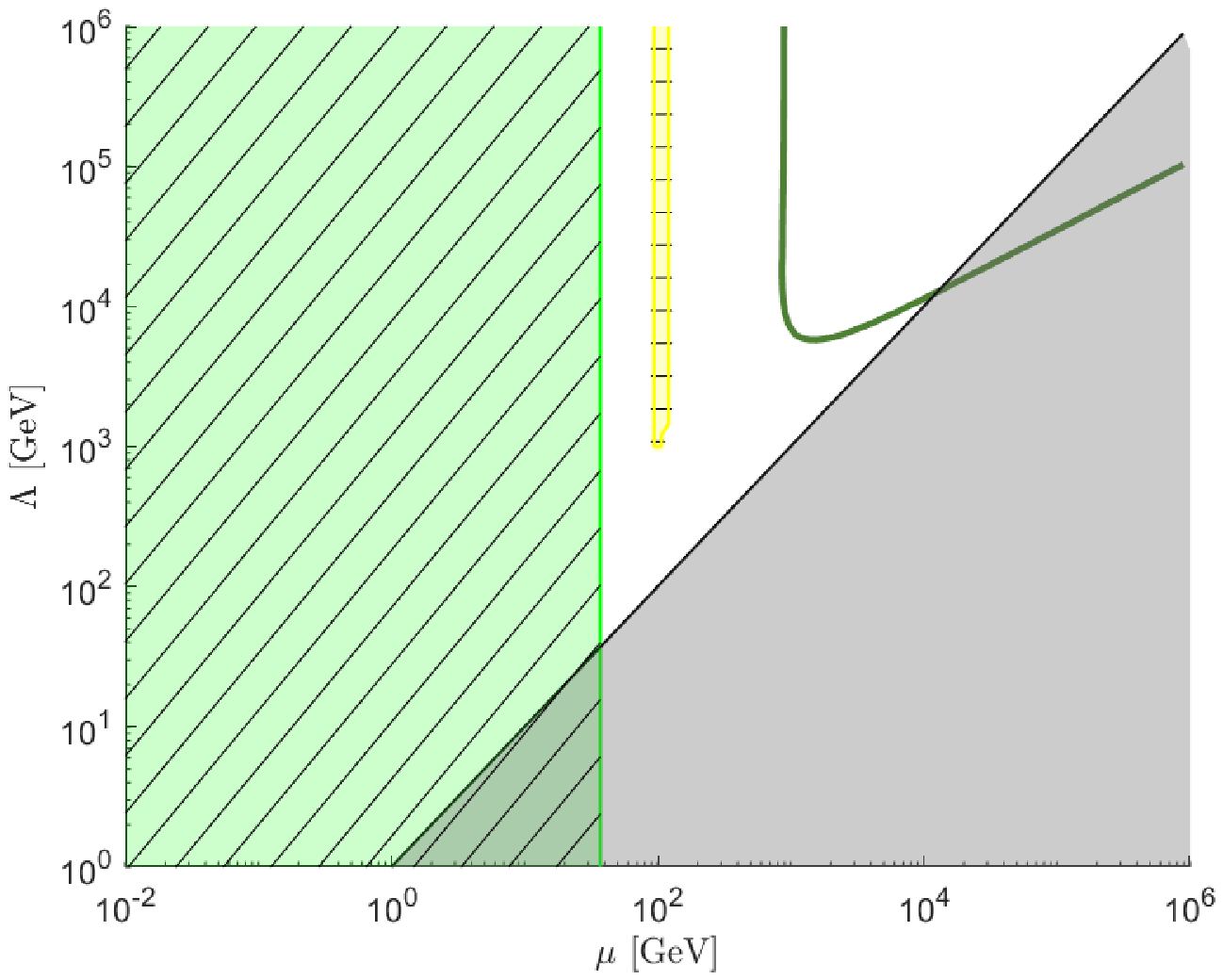}}
			\subfloat[$\frac{1}{\Lambda}\mathcal{O}_6^H$ from eq.~(\ref{eq:OH6})\label{fig:cW}]{\includegraphics[width=0.49\textwidth]{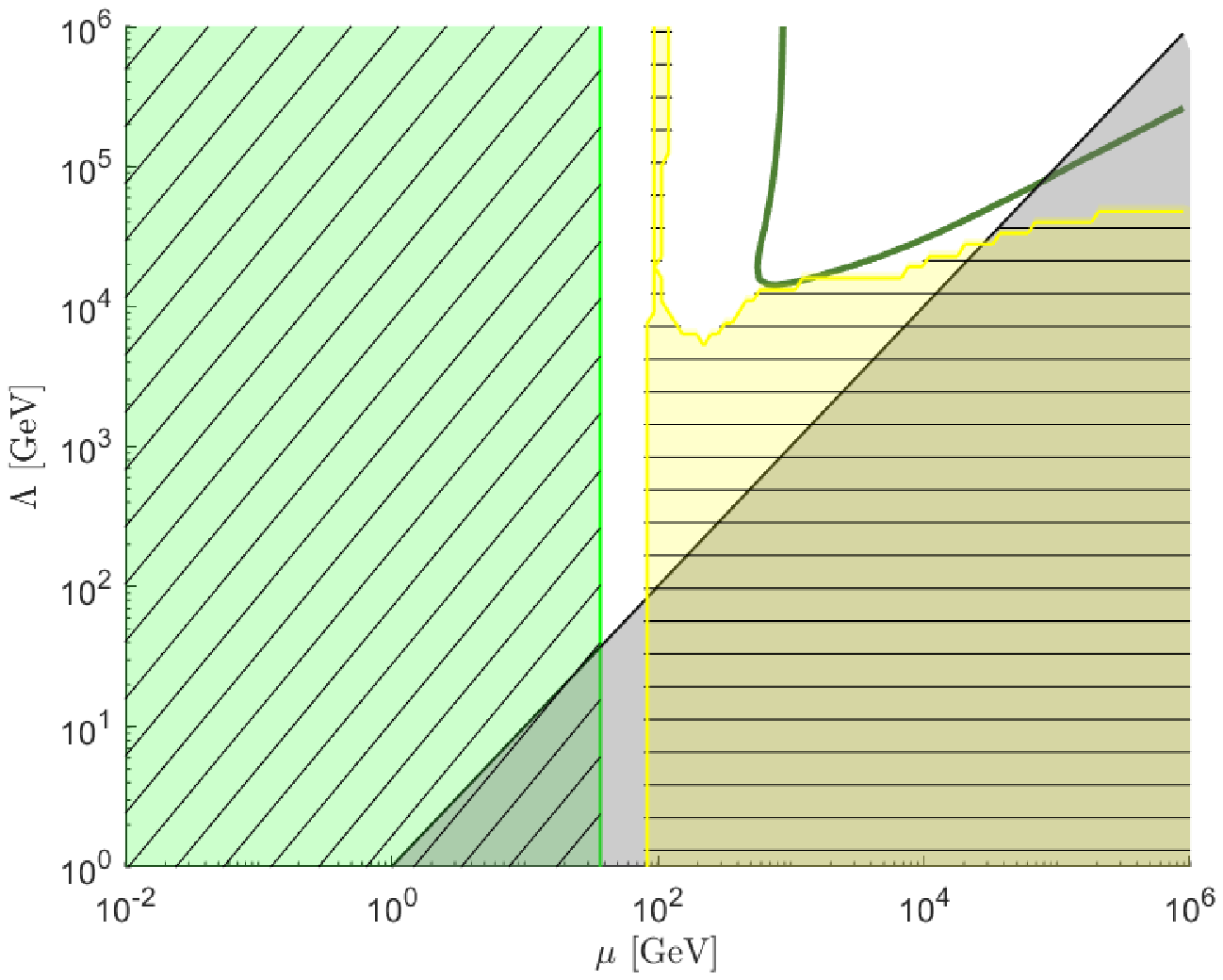}}
		}
		\caption{Parameter space for electric (top) and magnetic (bottom) dipole operators. The horizontal axes show the mass parameter $\mu$ and the vertical axes UV-cutoff allowed for the parameter point, determined with the numerical coefficients $c_i = 1$. The contours of parameter values which satisfy the relic abundance are shown as a solid green line. The regions where the EFT is not valid are shaded grey. The regions excluded by indirect detection are shaded yellow with horizontal hatching. The regions excluded by LEP are shaded  green with upwards diagonal hatching. \label{fig:higgsinodipole}}
	\end{figure}

	The parameter space for the electric dipole operators in eqs.~(\ref{eq:OH7}-\ref{eq:OH8}) are shown in Figs.~\ref{fig:cBCP}-\ref{fig:cWCP}. Both of these operators are CP-violating. Perhaps unsurprisingly, the effect of these CP-violating operators on the relic density and other experimental constraints is negligible. Rather, all of the relevant physics is determined by the dimension-4 terms in the Lagrangian, namely the interaction in eq.~(\ref{eq:higgsinodim4}).  Figures~\ref{fig:cBCP}-\ref{fig:cWCP} thus essentially represent the null hypothesis, where there is no impact from the higher dimension operators, which may be contrasted with the effects of the other operators. Hence, we may highlight the features of the parameter space before the inclusion of more interesting dimension-5 operators. Note that for these operators, the mass of the neutralinos and chargino is given by $\mu$. The lower right of the parameter space (small $\Lambda$, large $\mu$) represents the region where the effective field theory is not valid, as the physics occurs above the UV-cutoff. The region is bounded by the line $\mu = \Lambda$, which is the largest possible mass with a valid EFT.
	
	The observed relic density occurs at $\mu \simeq 1 \TeV$. For larger masses, the relic density increases and the universe becomes over-closed without some additional non-thermal mechanism to reduce the abundance. For smaller masses, the relic density decreases, apart from a resonance near the gauge boson masses. Also near the gauge boson masses, the resonance annihilation leads to indirect detection constraints. Lower masses are then subject to the LEP constraints on both the neutralino and chargino masses.

	\subsection{Higgsino magnetic dipole interaction via $W$}
	
	Figures~\ref{fig:cB}-\ref{fig:cW} show the parameter space for the operators in eqs.~(\ref{eq:OH5}) and (\ref{eq:OH6}) respectively. Together, these operators represent the magnetic dipole interaction of the higgsino. These operators provide point annihilations directly from two higgsinos to two gauge bosons as well as derivative interactions to a single gauge boson. The interactions increase in strength for decreasing $\Lambda$. The annihilation processes to a region of the parameter space for $\Lambda < 10\TeV$ and $\mu > m_W$ where the magnetic dipole interactions dominate the regular annihilation processes. The increase in the annihilation cross section in this region depletes the otherwise over-abundant dark matter, allowing for the relic density to be satisfied for masses greater than $1 \TeV$. The region allows for thermal relic higgsinos to have masses up to around $70 \TeV$, which corresponds to a magnetic dipole on the order of $10^{-3}$ to $10^{-4}$ times the proton magnetic moment.
	
	The operators also can be seen through indirect detection experiments, which can look directly for the annihilation products from the magnetic dipole. The operator $\mathcal{O}_6^H$ allows for direct annihilation into a $W^+W^-$ pair, whereas the $\mathcal{O}_5^H$ operator can only annihilate into a combination of $Z$ bosons and photons. The indirect detection bounds are given for the $W^+W^-$ channel only~\cite{fermilat14}. Meanwhile, the reduced abundances due to the additional annihilation reduce the total amount of annihilation in the $W^+W^-$ channel.

	\subsection{Higgs-higgsino interaction with neutralino mass splitting}
	\begin{figure}[t]
	\centering
	{
		\subfloat[Mass contour for $\frac{c_{1,2}}{\Lambda}(\mathcal{O}_1^H + \mathcal{O}_2^H)$ with $c_{1,2} = -1$ (top) and $c_{1,2} = +1 $ (bottom) \label{fig:Mc1}]{
			\begin{tabular}[c]{c}
			\includegraphics[width=0.4\textwidth]{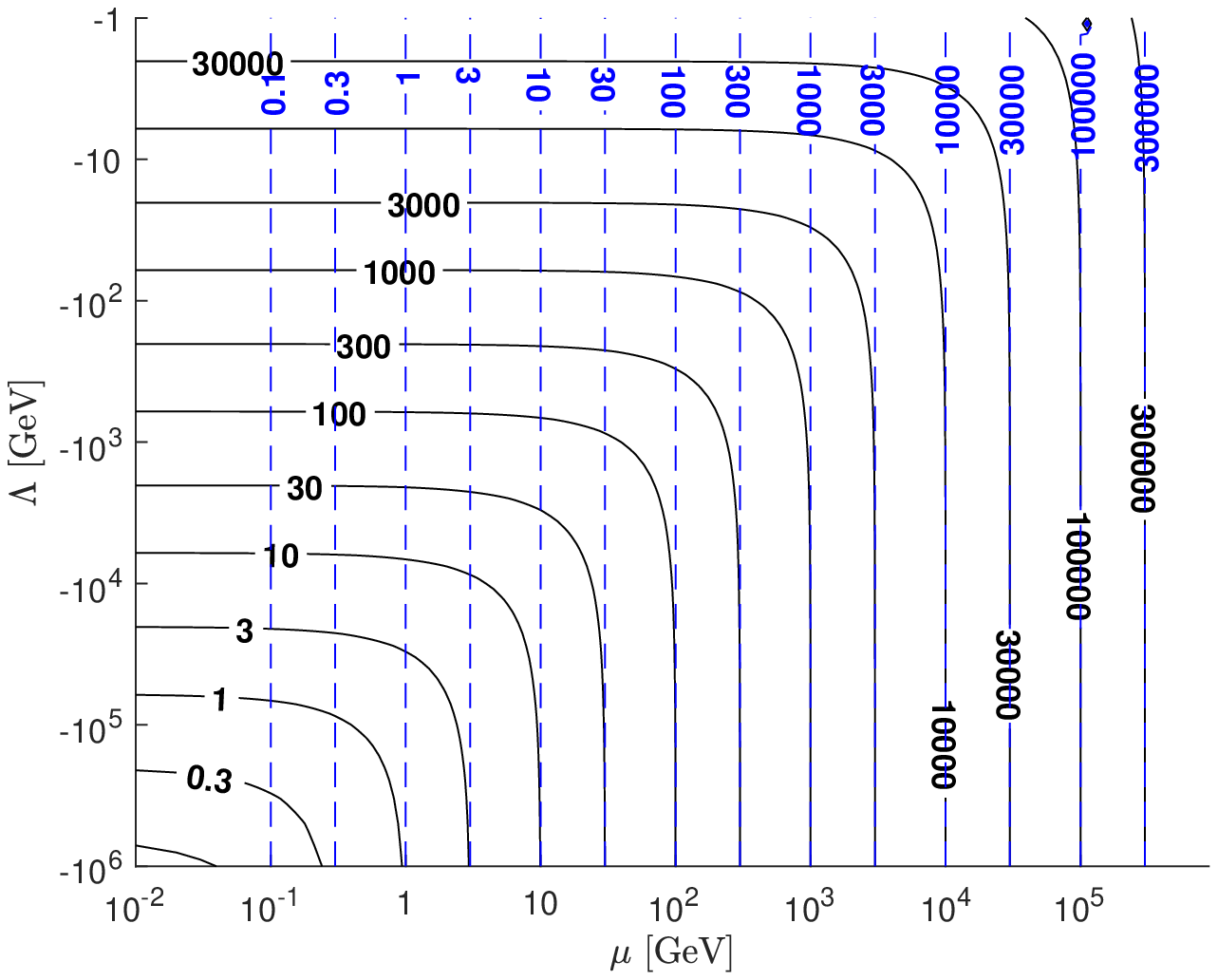}
			\\
			\includegraphics[width=0.4\textwidth]{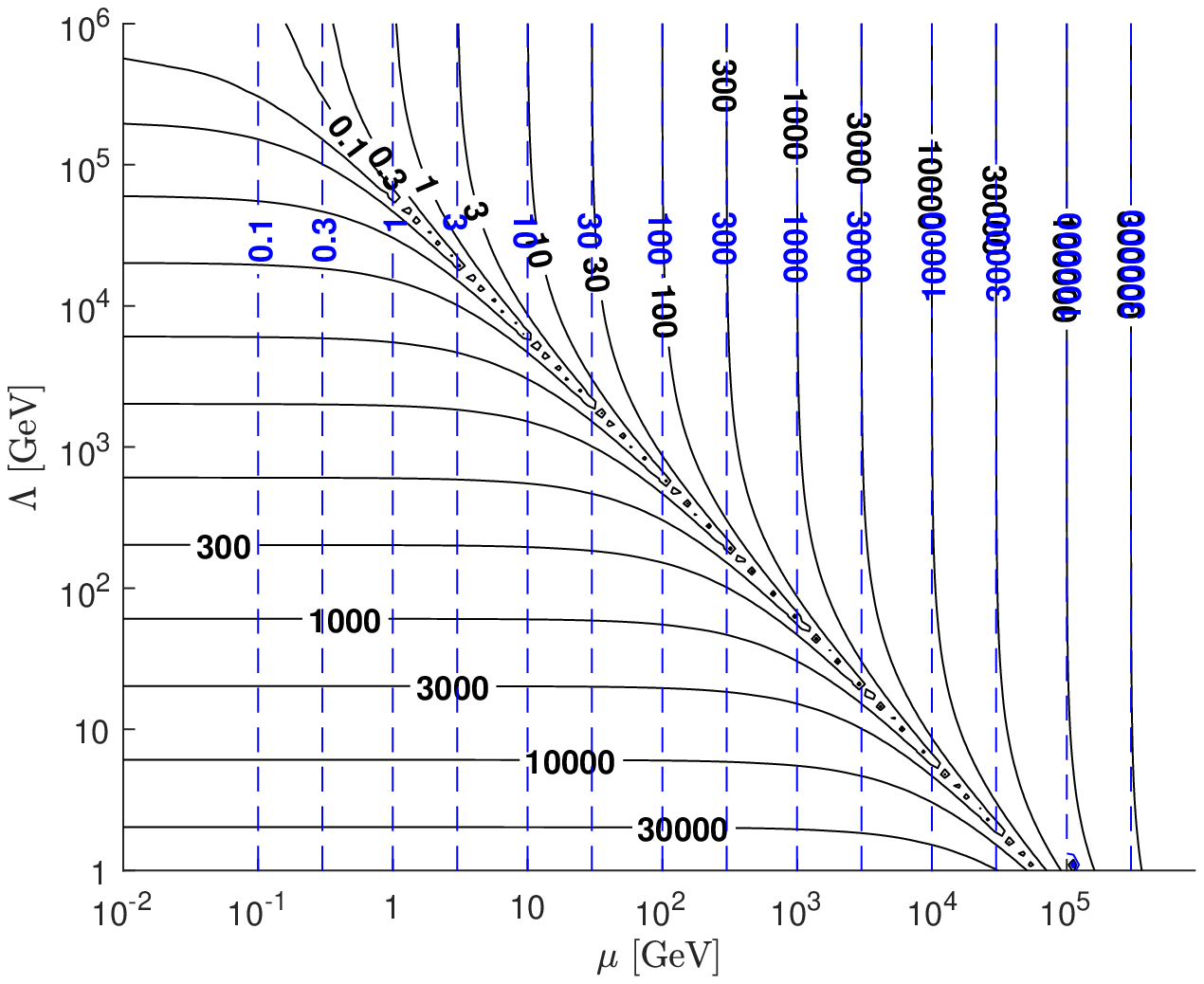}
			\end{tabular}
		}
		\subfloat[$\frac{c_{1,2}}{\Lambda}(\mathcal{O}_1^H + \mathcal{O}_2^H)$ from eqs.~(\ref{eq:OH1}-\ref{eq:OH2}) with $c_{1,2} = -1 $ (top) and $c_{1,2} = +1$ (bottom)\label{fig:c1}]{
			\begin{tabular}[c]{c}
			\includegraphics[width=0.45\textwidth]{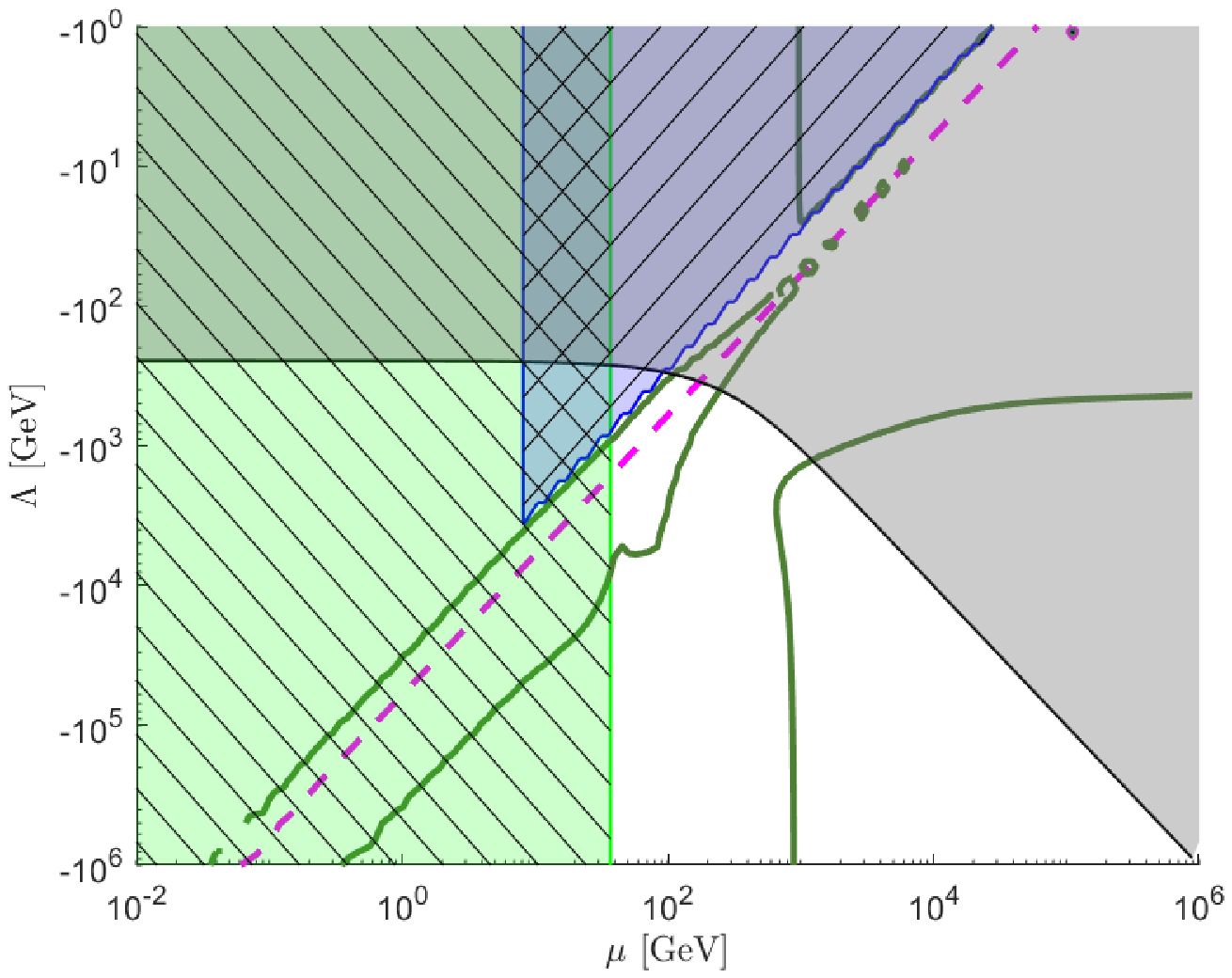}
			\\
			\includegraphics[width=0.45\textwidth]{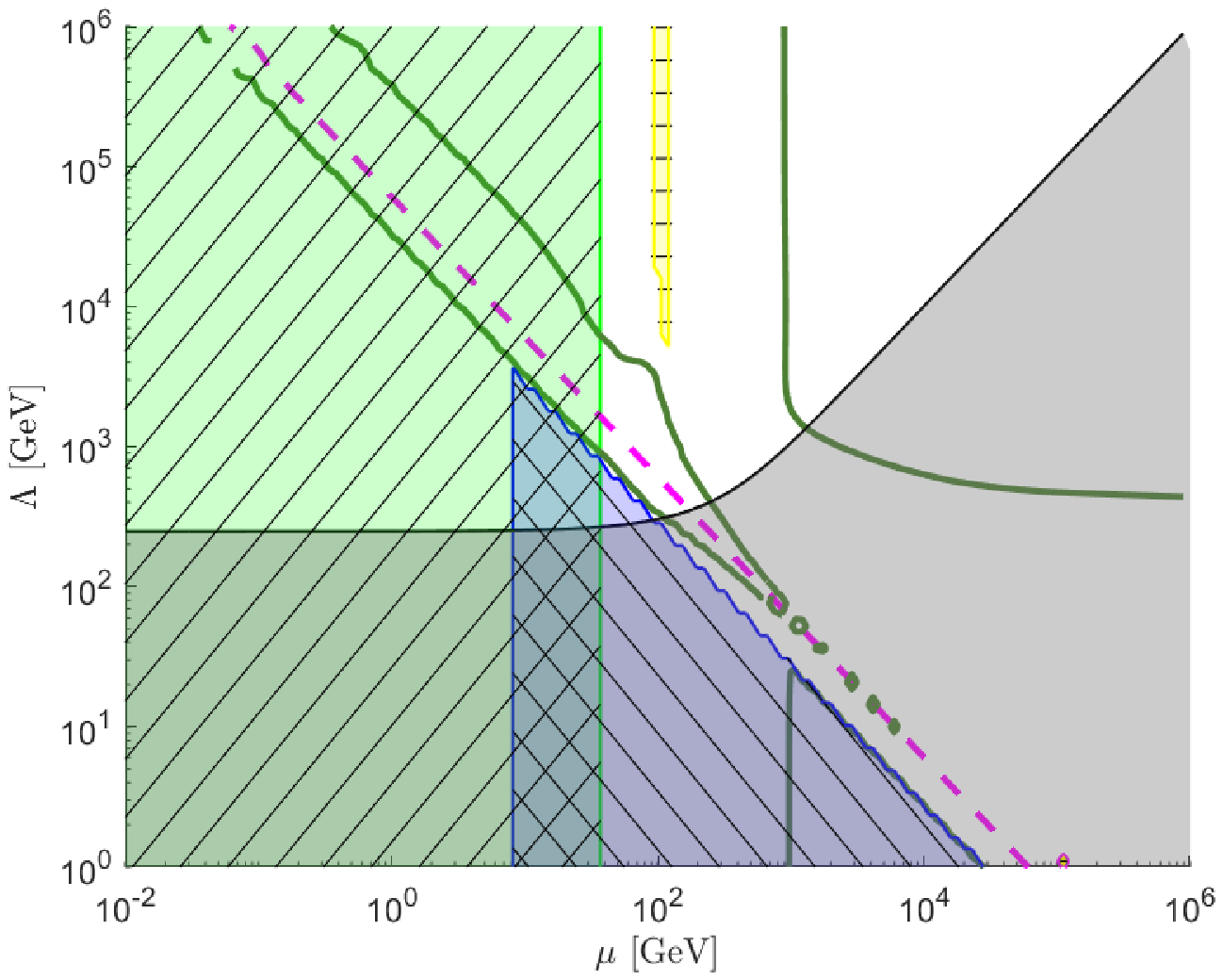} 
			\end{tabular}
		}
	}
	\caption{Mass contour plots (left) and parameter space (right) for Higgs-higgsino operators. Mass contour shows the lightest neutralino mass in black and the lightest chargino mass in blue as a function of the input parameters. The top graphs show negative values of the coupling constant and the bottom graphs show the positive values. The region where $\mu \simeq \frac{v^2}{\Lambda}$ is marked as a dashed magenta line. The regions excluded by direct detection are shaded blue with downwards diagonal hatching. For a description of the parameter space plots, see the caption of Fig.~\ref{fig:higgsinodipole}.\label{fig:higgsinohiggs12}}
	\end{figure}
	\begin{figure}[t]
	\centering
	{	
		\subfloat[Mass contour for $\frac{c_{3,4}}{\Lambda}(\mathcal{O}_3^H + \mathcal{O}_4^H)$ for $c_{3,4} = -1$ (top) and $c_{3,4} = +1$ (bottom)\label{fig:Mc3}]{
			\begin{tabular}[c]{c}
			\includegraphics[width=0.4\textwidth]{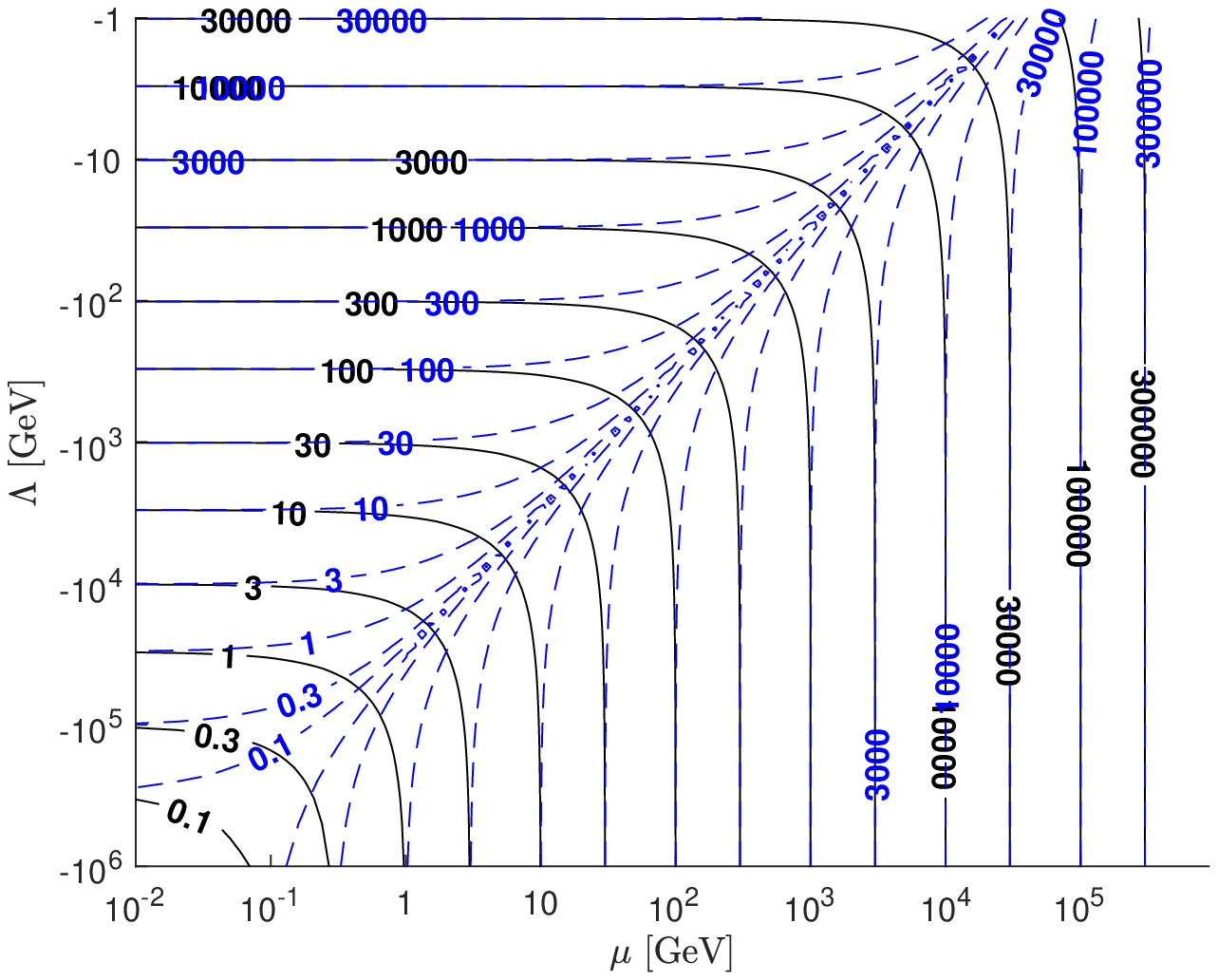}
			\\
			\includegraphics[width=0.4\textwidth]{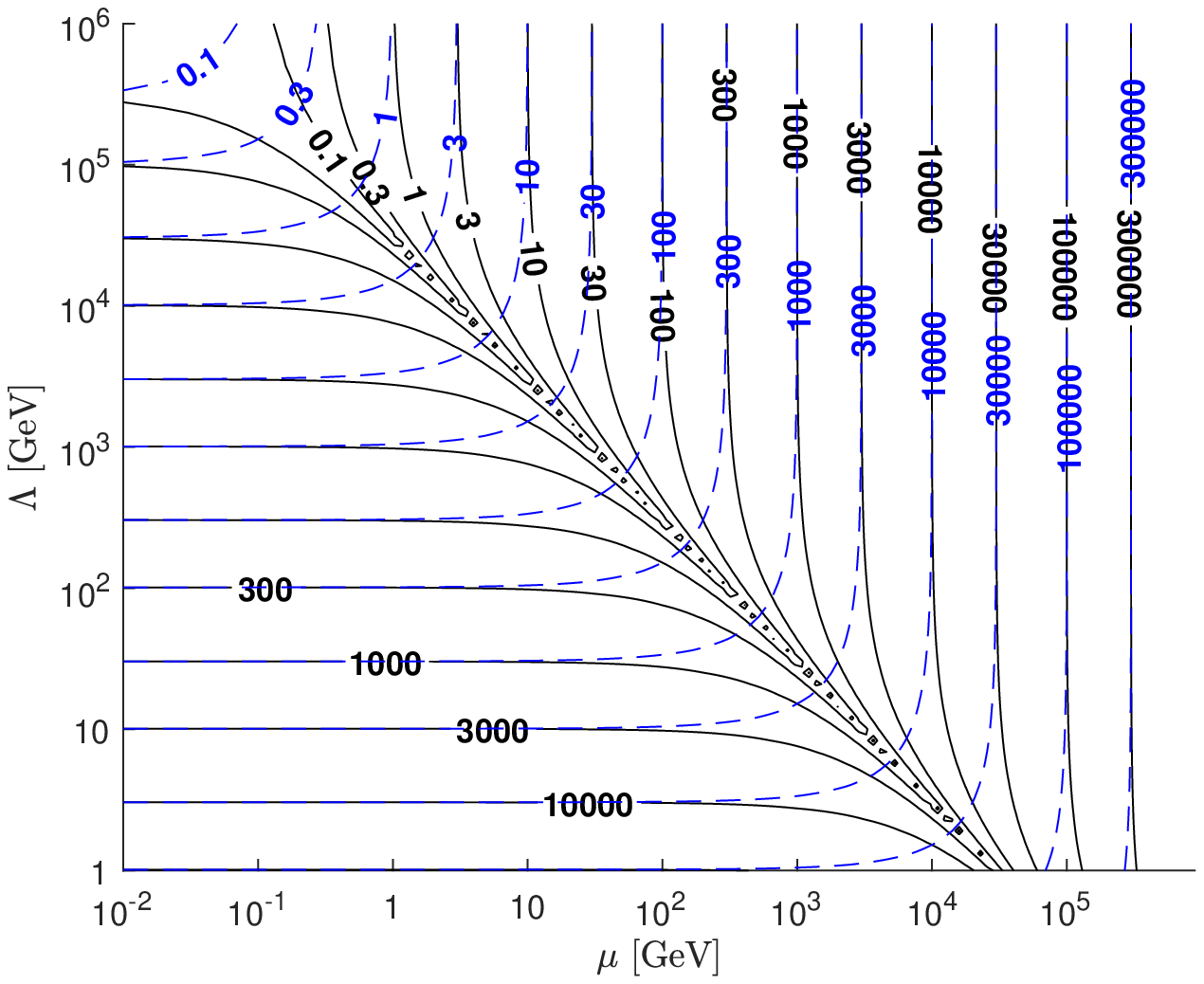}
			\end{tabular}
		}
		\subfloat[$\frac{c_{3,4}}{\Lambda}(\mathcal{O}_3^H + \mathcal{O}_4^H)$ from eqs.~(\ref{eq:OH3}-\ref{eq:OH4}) for $c_{3,4} = -1$ (top) and $c_{3,4} = +1$ (bottom)\label{fig:c3}]{
			\begin{tabular}[c]{c}
			\includegraphics[width=0.45\textwidth]{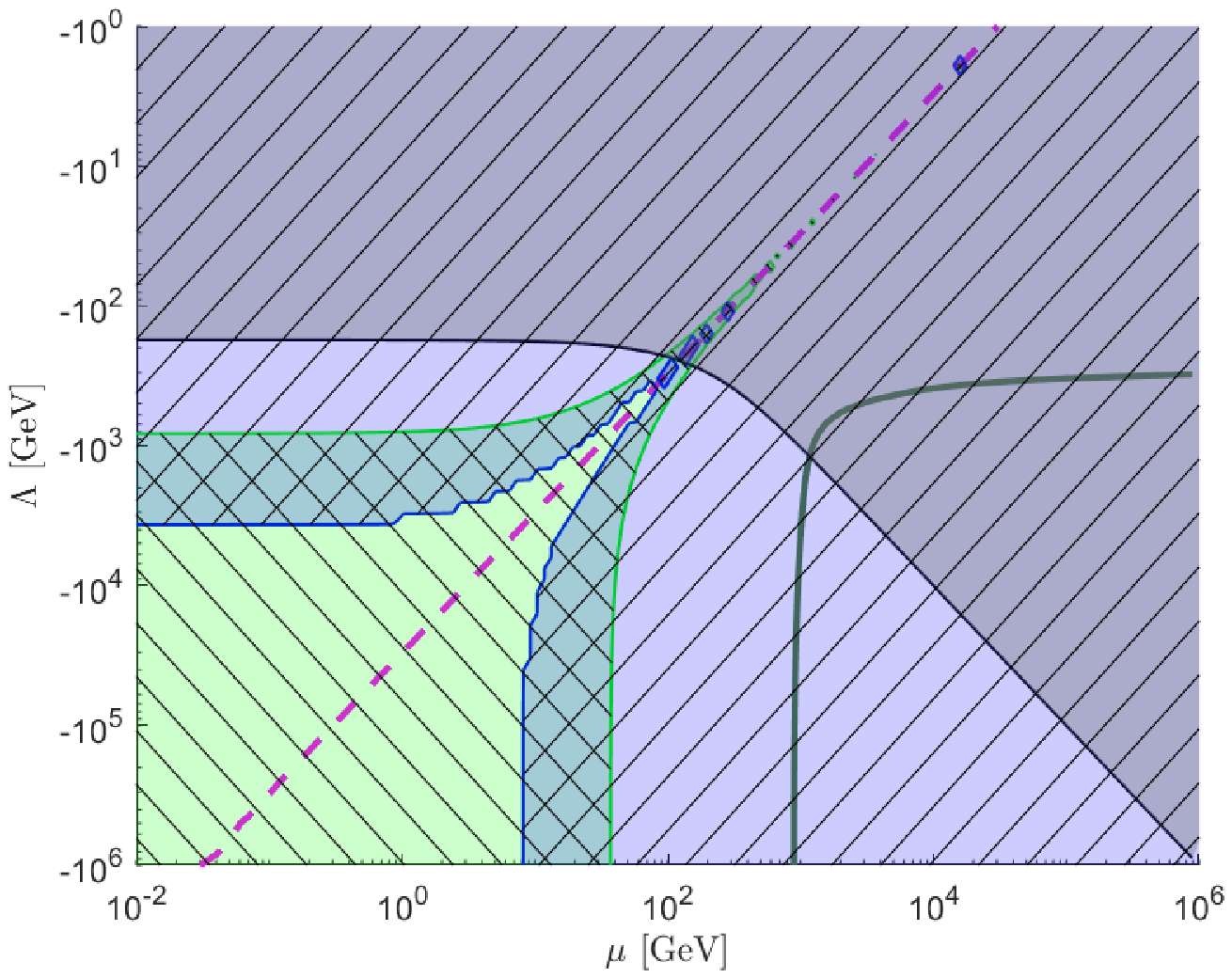}
			\\
			\includegraphics[width=0.45\textwidth]{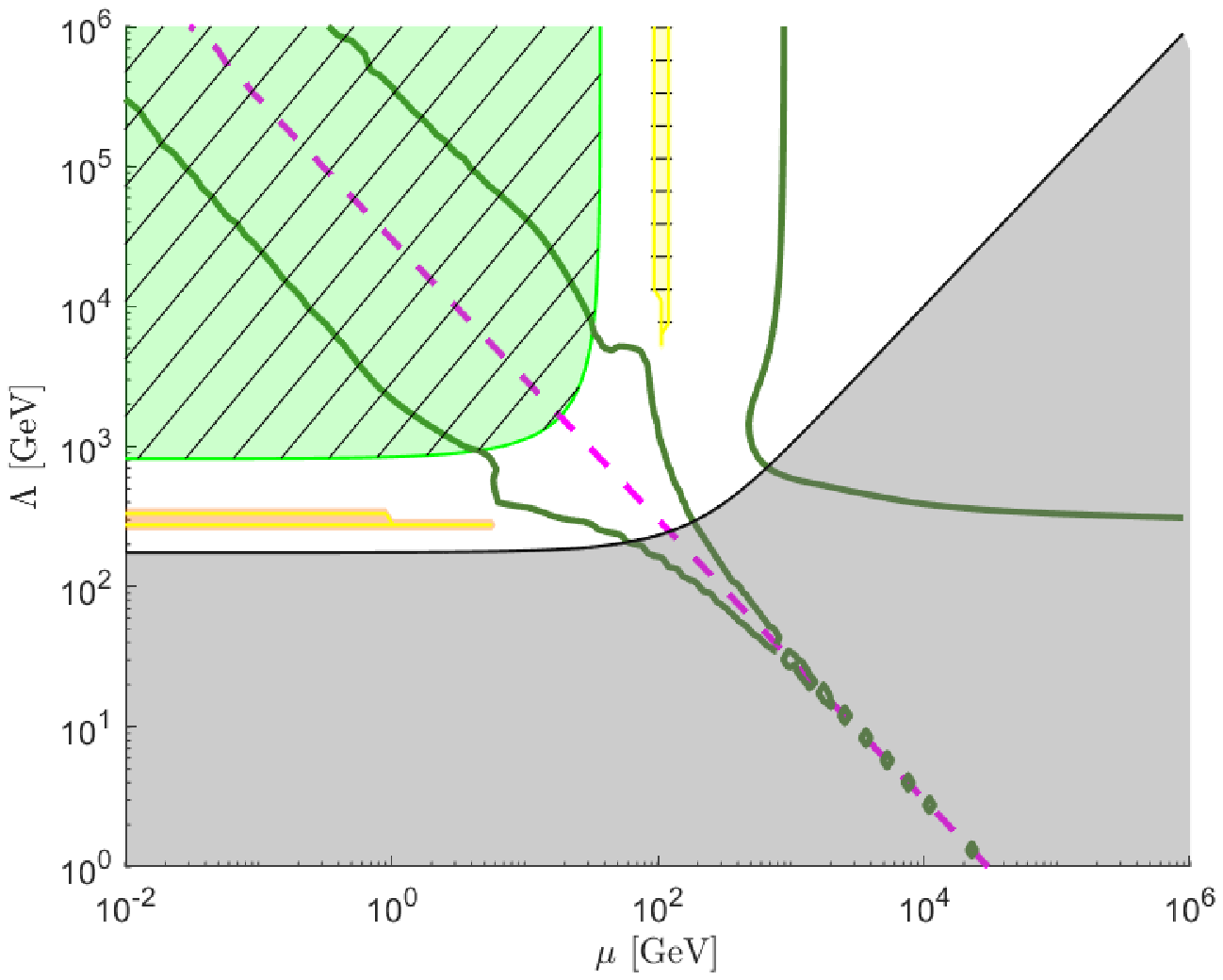}
			\end{tabular}
		}
	}
	\caption{Mass contour plots (left) and parameter space (right) for Higgs-higgsino operators. Mass contour shows the lightest neutralino mass in black and the lightest chargino mass in blue as a function of the input parameters. The regions excluded by direct detection are shaded blue with downwards diagonal hatching.  For a description of the parameter space plots, see the caption of Fig.~\ref{fig:higgsinodipole}.\label{fig:higgsinohiggs34}}
\end{figure}
	
	The Higgs-higgsino interaction terms provide significantly different phenomena than the higgsino dipole interactions. Most notably, the dimension-5 terms will contribute mass terms for the higgsino after the Higgs Boson undergoes symmetry breaking. The mass of the two higgsinos and chargino become non-degenerate. Figures~\ref{fig:Mc1} and \ref{fig:c1} show the parameter space for the Lagrangian terms $\frac{c_{1,2}}{\Lambda}(\mathcal{O}_1^H + \mathcal{O}_2^H)$ from eqs.~(\ref{eq:OH1}-\ref{eq:OH2}) where, for simplicity, we have assumed $c_1 = c_2 = c_{1,2}=\pm1$. These terms are combined as they provide for a mass splitting between the two neutralinos through eqs.~(\ref{eq:m1}-\ref{eq:m2}). The consequence of the two terms being equal is that we do not have an offset to $\bar{\mu}$ per eq.~(\ref{eq:mutilde}), which is dependent on the difference between the two coefficients. The mass of the chargino does not depend on $c_1$ or $c_2$, and so is given by $\mu$.
	
	For $\mu \gg \frac{v^2}{\Lambda}$, that is, $\mu$ and $\Lambda$ are both large, all of the masses are degenerate and equal to $\mu$. 
	For $\mu \ll \frac{v^2}{\Lambda}$, that is, $\mu$ and $\Lambda$ are both small, the higgsino masses are both given by $\frac{c_{1,2} v^2}{\Lambda}$, but the chargino mass is still $\mu$ and there is a sign difference between the two neutralinos. There is therefore a mass splitting between the chargino and the higgsinos. Notably, in this region, the chargino becomes the lightest supersymmetric particle, instead of the higgsino. The region is strongly ruled out as it interacts too strongly with the Standard Model through the presence of the electric charge.
	
	In the regime that $\mu \sim \frac{v^2}{\Lambda}$, shown in Figures~\ref{fig:c1} as a dashed magenta line, the mass of the lightest higgsino approaches zero, while the mass of the chargino remains at $\mu$. The heaviest higgsino is now twice as massive as the chargino. Hence, the degeneracy between the chargino and the neutralino is broken. There is a significant impact on the relic density as the dominant annihilation mode is the coannihilation mode where near-degenerate higgsinos and charginos annihilate into fermions via a $W$ boson. Indeed, in the limit that the winos and binos (if present) have heavy mass, that is, the two neutralinos are an identical mix of the two higgsinos, the annihilation between neutral higgsinos vanishes~\cite{mcdonald92}. This is because of a cancellation between the two higgsino-higgsino-$Z$ boson terms in the Lagrangian. In the canonical case, the annihilation via the chargino and $W$-boson exchange is dominant~\cite{mizuta93}. Hence, by removing the mass degeneracy we no longer have a viable annihilation processes, leaving the universe to be overclosed by higgsino dark matter. There is then a continuous variation away from the case where $\mu \sim \frac{v^2}{\Lambda}$ where the coannihilation pathway becomes unlocked. Along this variation is a point whereby the annihilation rate corresponds to the relic density. The dark matter mass along this contour ranges from $m_\chi \sim 50 \GeV$ to $m_\chi \sim 100\GeV$. The lower bound arises from the LEP bounds, whereas the upper bound is limited by the range of the EFT.
	
	\subsection{Higgs-higgsino interaction without neutralino mass splitting}
	Figures~\ref{fig:Mc3} and \ref{fig:c3} show the mass contour and parameter space for the operators which do not split the neutralino masses $\frac{c_{3,4}}{\Lambda}(\mathcal{O}_3^H + \mathcal{O}_4^H)$ from eqs.~(\ref{eq:OH3}-\ref{eq:OH4}). The parameter space is similar to the mass splitting operators above, however the two neutralinos are still degenerate throughout as the $c_3$ amounts to a shift in the $\mu$ components in the mass matrix. Now, the chargino mass is also affected by the $c_4$ component in eq.~(\ref{eq:mpm}), which is disconnected from the modified $\mu$ parameter in eq.~(\ref{eq:mubar}).
	
	For $\mu \gg \frac{v^2}{2\Lambda}$, that is, both $\mu$ and $\Lambda$ are large, the masses of both the higgsinos and charginos are degenerate and equal to $\mu$. There is no impact from the dimension 5 operators in this regime.
	For $\mu \ll \frac{v^2}{2\Lambda}$, that is, both $\mu$ and $\Lambda$ are small, the masses of the higgsinos are again both large and degenerate, with $m_1 = m_2 = \frac{v^2c_3}{2\Lambda}$ and $m_\pm = \frac{v^2c_4}{2\Lambda}$. As we have set $c_3 = c_4$, the chargino and neutralino are also still degenerate.
	
	Note that the masses in the parameter space are symmetric about $\mu = \frac{v^2}{2\Lambda}$. The indirect detection bounds and relic density bounds are therefore also reflected about the line. Along the line $\mu \sim \frac{v^2}{\Lambda}$, shown as a dashed magenta line in Figure \ref{fig:c3}, the masses of neutralinos approach zero but the chargino mass is doubled, as per the mass splitting case. Again, the annihilation processes that contribute the most to the higgsino relic density are the coannihilations with the chargino, for example the $t$-channel annihilation to $W$-bosons. Where there is the near cancellation between the two terms, the coannihilation processes are no longer efficient and so the annihilation does not proceed. Like the mass splitting case, there is a continuous variation between the highly fine-tuned cancellation and the standard case, along which there is a set of parameters where the relic density is met. There is a viable region for relic higgsino dark matter with masses $m_\chi = 80 \GeV$ to $m_\chi = 120 \GeV$. The mass cancellation also affects the relic density near $1 \TeV$, reducing the allowed region as low as $m_\chi = 500 \GeV$. Finally, below $\Lambda = 10^3 \GeV$, the direct annihilation from the additional terms becomes signficant and dominates the annihilations, but this only corresponds to small region of the viable parameter space, because for $\Lambda < v$ the effective field theory breaks down. 
	
	However, when the coupling parameter is negative, the mass cancellation now occurs in the chargino rather than the neutralino as per eq.~(\ref{eq:mpm}). In the valley where $\mu \sim \frac{v^2}{\Lambda}$, there is no additional regime with a viable neutralino relic density, and the parameter space is greatly restricted by the direct detection constraints on chargino dark matter. As the argument of the complex coupling constant transitions from $0$ to $\pi$, there is a continuous transition from one scenario to the other, with an inflection point at $\frac{\pi}{2}$ where the lightest particle transitions from the neutralino to the chargino.

	\subsection{Wino inelastic magnetic dipole}

	\begin{figure}[t]
	\centering
	{
	
		\subfloat[$\frac{1}{\Lambda}\mathcal{O}_2^W$ from eq.~(\ref{eq:OW3})]{\label{fig:winomagdipinel}\centering\includegraphics[width=0.49\textwidth]{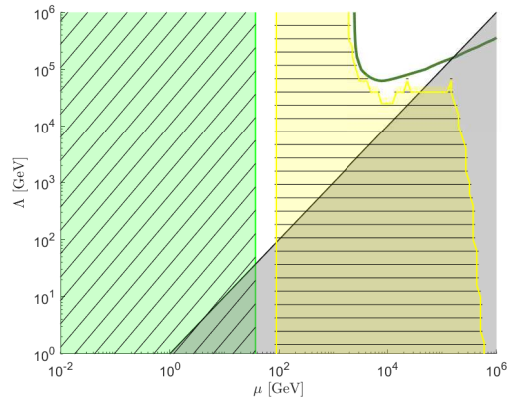}}
	}
	\caption{Parameter space for wino magnetic dipole operator. For a description, see the caption of fig.~\ref{fig:higgsinodipole}}
	\end{figure}
	
	The wino inelastic dipole with Lagrangian in eq.~(\ref{eq:OW3}) is shown in Figure \ref{fig:winomagdipinel}.
	The results and physics are similar to the higgsino case. When the cut-off is large, the dipole component has a negligible effect, and the resultant relic density occurs at a mass equal to the canonical $3\TeV$. However, such relic densities are ruled out by indirect detection, magnified due to the Sommerfeld enhancement~\cite{slatyer10,feng10,cassel10}. Again, where the dipole is large, there is additional annihilation in the early universe which can reduce the overabundance of the dark matter density for larger masses. The additional annihilation allows for thermal relics with masses up to $\sim 100 \TeV$ in the case that $d_3 = 1$.
	The addition of the dipole also allows for indirect detection through the annihilation diagrams arising directly from the dipole. There is therefore a region of the parameter space below $\Lambda \sim 3 \TeV$, but above the $W$ and $Z$ masses where the model is excluded by the indirect detection constraints.
	\subsection{Wino-Higgs interaction}

	\begin{figure}[t]
	\centering
	{	
	\subfloat[Mass contour for $\frac{d_{1}}{\Lambda}\mathcal{O}_W^1$\label{fig:masswinohiggsnosplitting} for $d_1 = -1$ (top) and $d_1 = +1$ (bottom)]{
		\begin{tabular}[c]{c}
		\includegraphics[width=0.4\textwidth]{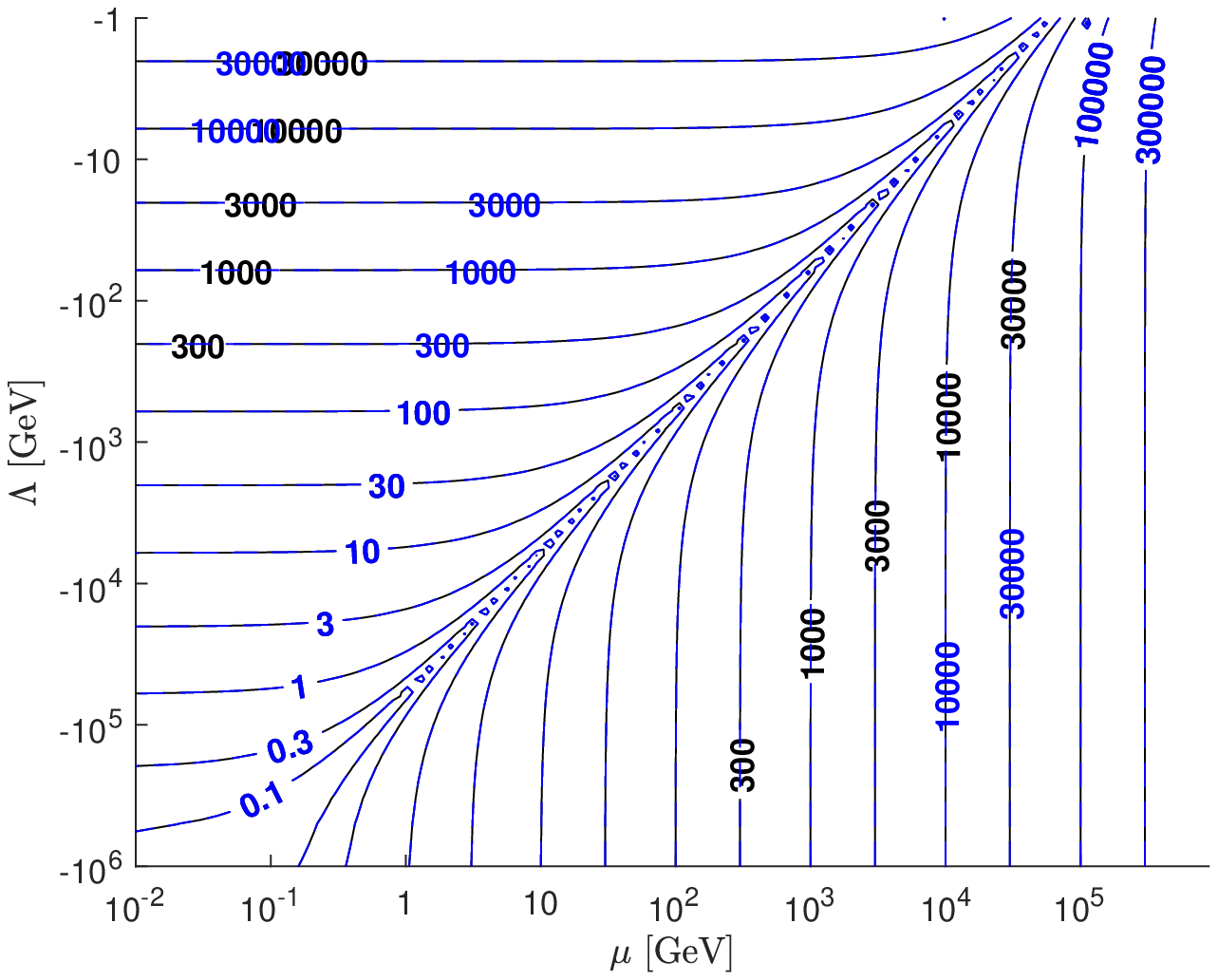}
		\\
		\includegraphics[width=0.4\textwidth]{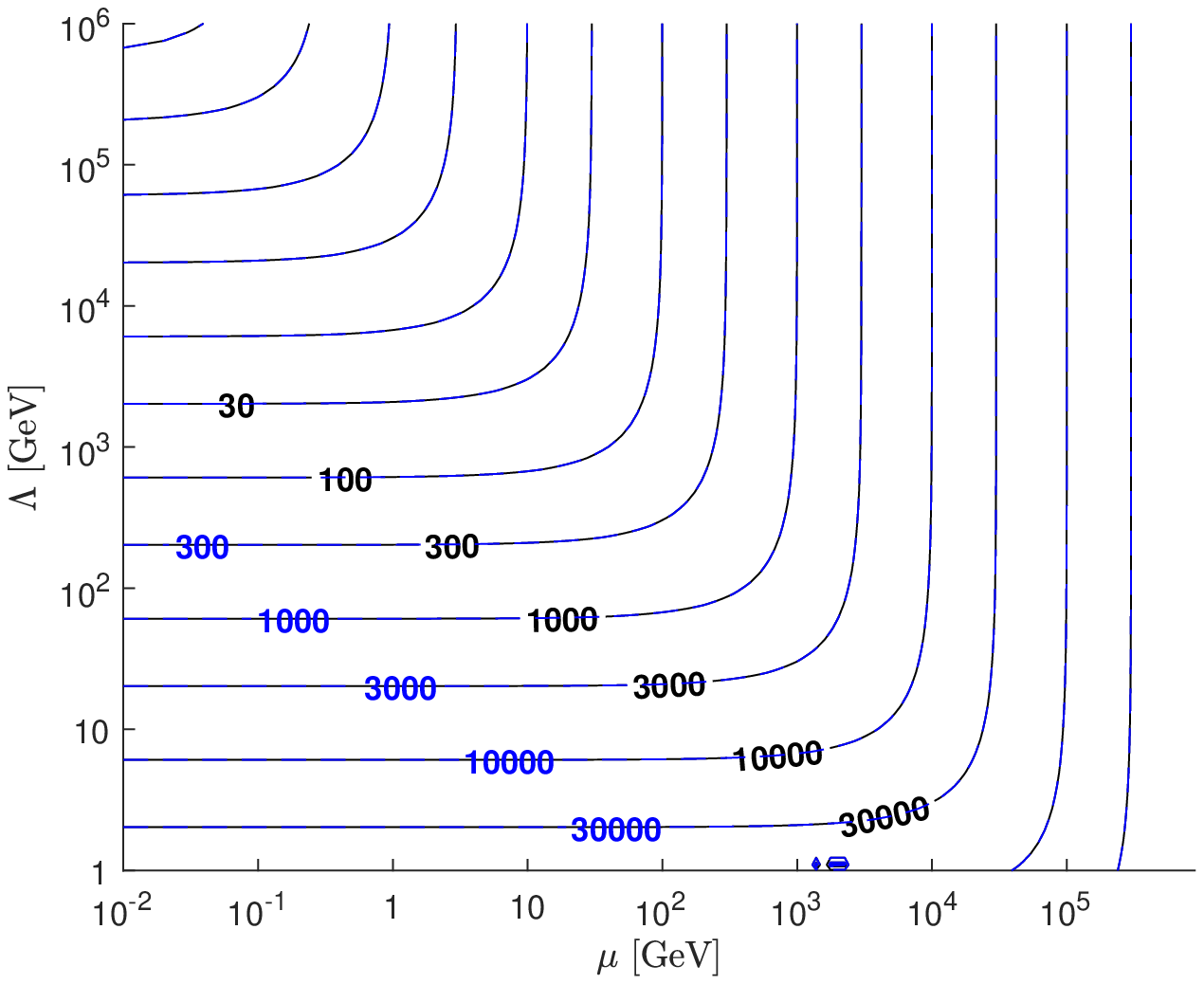}
		\end{tabular}
		}
	}{
	\subfloat[$\frac{d_{1}}{\Lambda}\mathcal{O}_W^1$ from eq.~(\ref{eq:winohiggsnosplitting}) for $d_1 = -1$ (top) and $d_1 = +1$ (bottom)]{\label{fig:winohiggsnosplitting}
		\begin{tabular}[c]{c}
		\includegraphics[width=0.45\textwidth]{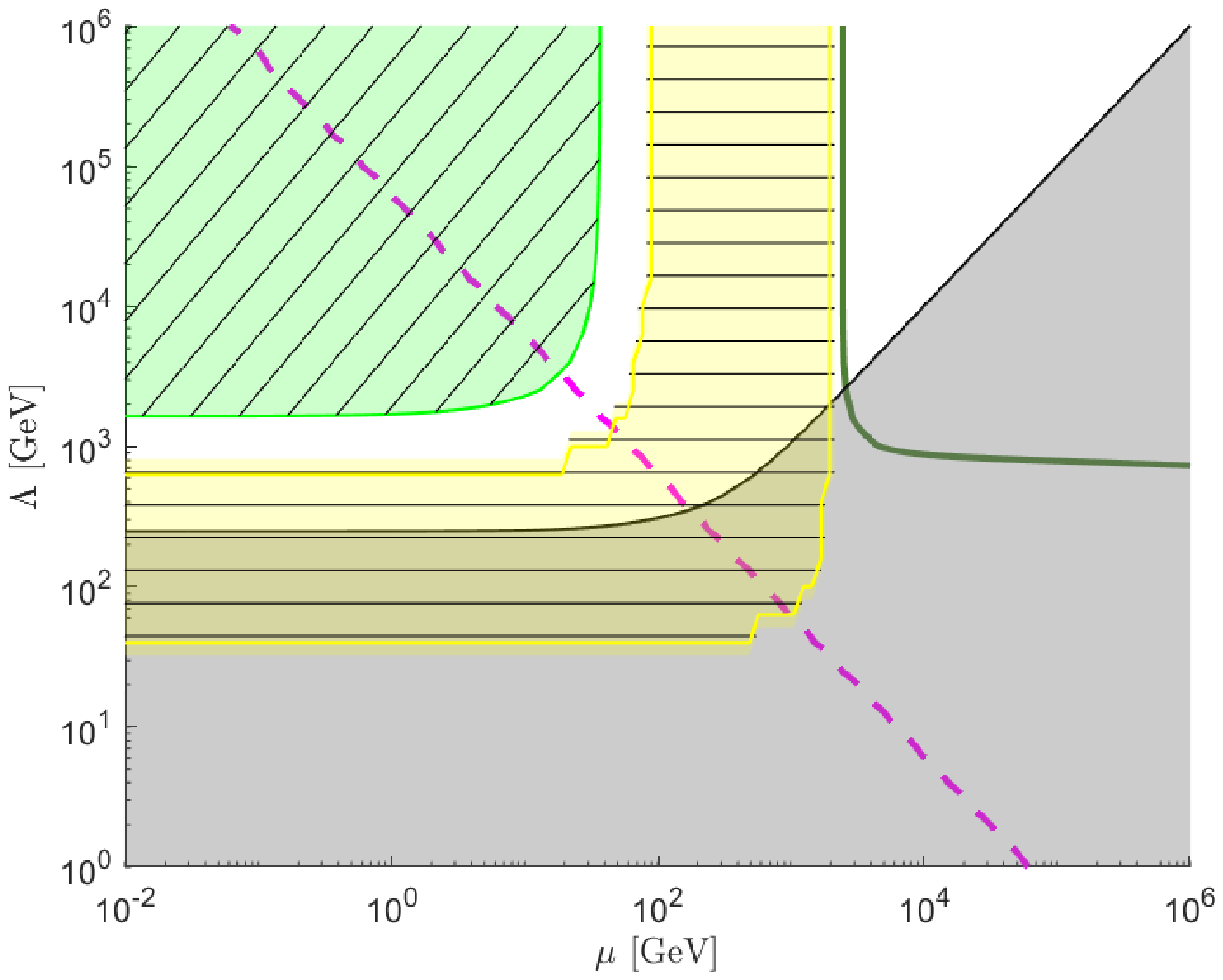}
		\\
		\includegraphics[width=0.45\textwidth]{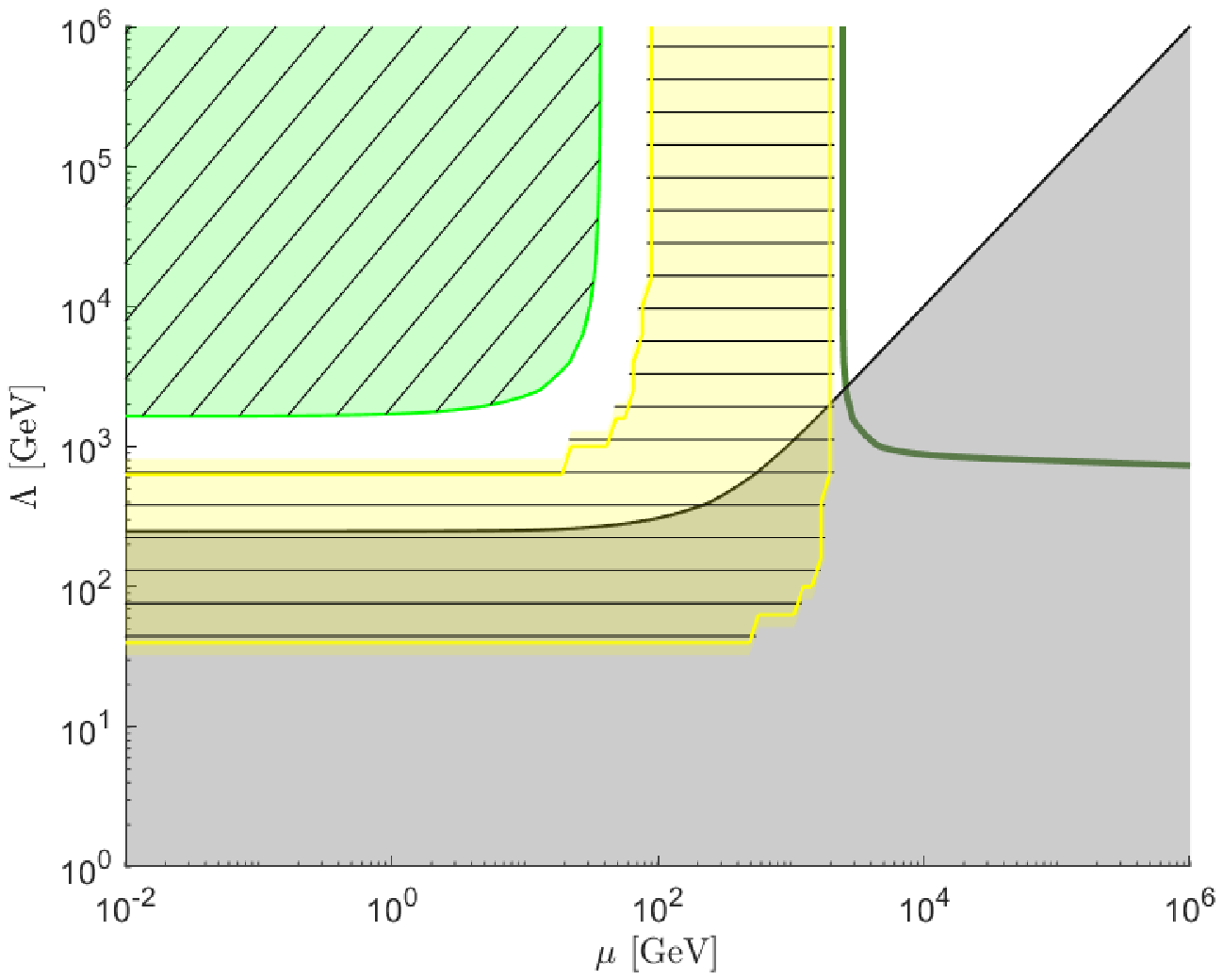}
		\end{tabular}
		}	
	}

	\caption{Mass contour plots (left) and parameter space (right) for wino-Higgs operators. For a description of the mass contour plots, see the caption of fig.~\ref{fig:higgsinohiggs12}. For a description of the parameter space plots, see the caption of fig.~\ref{fig:higgsinodipole}. }
	\end{figure}

	Figures~\ref{fig:masswinohiggsnosplitting} and ~\ref{fig:winohiggsnosplitting} shows the mass spectrum and parameter space of the operators in eqs.~(\ref{eq:winohiggsnosplitting}\winobino{-\ref{eq:binohiggsnosplitting}}) which shift the mass of the wino. Like in the higgsino cases, for $\frac{d_1v^2}{\Lambda} < M_2$, the mass term is dominated by the mass parameters in the Lagrangian and the effect of the dimension-5 term is negligible. For $\frac{(d_2+d_3)v^2}{\Lambda} > M_2$, the contributions from the dimension-5 terms dominate, and the relic density increases for decreasing $\Lambda$. Thus, the observed relic density follows the mass contour through the parameter space for $m_\chi = 3\TeV$. However, unlike in the higgsino case, we have an additive mass correction\winobino{ in eq.~(\ref{eq:mT})}, rather than a cancelling correction \winobino{like in eq.~(\ref{eq:m1})}, as well as a contribution to the chargino mass\winobino{ in eq.~(\ref{eq:mpmwino})}.  Hence, the behaviour of the relic density is completely determined by the measurable masses of the dark matter particles.
	
\winobino{	\subsection{Wino-Higgs Interaction with Mass Splitting}

	Figures~\ref{fig:masswinohiggsplitting} and \ref{fig:winohiggssplitting} show the mass spectrum and parameter space of the operator in eq.~(\ref{eq:winohiggssplitting}), which does form a splitting in the masses between the two neutralinos and between them and the chargino. Unlike the higgsino case, the mass of the chargino is unchanged by these operators because by combining the gaugino triplet and singlet, we do not obtain an overall neutral combination containing charged gauginos. 
	
	However, by eqs.~(\ref{eq:Wm1}-\ref{eq:Wm2}), there is either an additive or subtractive contribution to the average of the interaction basis gaugino masses. Once again, there is a mass cancellation in the case that $M_T = \sqrt{\Delta M^2 + \frac{c_1^2 v^4}{\Lambda^2}}$ with physics equivalent to the higgsino and wino cases, that is, the coannihilation processes are suppressed. If $\Delta M^2 + \frac{c_1^2 v^4}{\Lambda^2} > M_T^2$, then the lightest particle is the chargino, and the resultant model is not a valid description of dark matter. The physics is broadly equivalent to the mass splitting higgsino case as described above. However in this case, the UV-cutoff discounts any region where the direct annihilation is dominant, and the LEP bounds and indirect detection together exclude the region near the mass cancellation

}
	

%% file: results.tex
\section{Discussion}

	We have introduced a series of effective field theory operators for composite higgsino-like and wino-like dark matter. 

The operators contribute to both annihilation to Higgs bosons and gauge bosons, the latter having the form of an inelastic electromagnetic dipole. We have calculated the relic density for a range of values in the mass-coupling parameter space. We then compared the viability to experimental constraints, including direct and indirect detection experiments and LEP bounds. 
	
	Most of the operators considered have a region of the parameter space for sufficiently strong couplings where annihilation via the dimension-5 operators dominates the relic density computation. The additional annihilation reduces the relic density as the depletion of dark matter is more efficient. However, the operators which decay into a pair of Higgs bosons require coupling strengths which are too large to be accounted for by an effective field theory, as the interaction occurs at a scale well above the UV-cutoff of the theory. For the models which involve an electromagnetic dipole, inelastic magnetic dipoles with strength $\mathcal{O}(10^{-3})$ times the proton magnetic moment result in a viable annihilation pathway within the bounds of the effective field theory. These models allow for viable thermal relic higgsino-like or wino-like dark matter with masses up to $\sim 10\TeV$ or $\sim 100\TeV$ respectively. Higher mass models are not valid in the present effective field theory description. These regions will be detectable through indirect detection experiments with a $\mathcal{O}(10)$ increase in the sensitivity of indirect detection experiments. 
	
	We have also seen that the operators which couple to a Higgs boson can produce shifts in the masses of the neutralinos once the Higgs bosons have undergone electroweak symmetry breaking. The operators produces mass splittings between the neutralinos and between the neutralinos and charginos. These splittings cause the lightest neutralino mass to approach zero due to a cancellation in the mass term between the given mass and the dimension-5 coupling. As the mass decreases, the loss of degeneracy to the charginos reduces the efficiency of the coannihilation processes. There is then a region of the parameter space which is a viable thermal relic, with a masses near $100\GeV$. These regions are bound by the LEP constraints from the invisible $Z$-width, and may be within range of near-future experiments. The required strength of these operators is too big to be produced by the conventional MSSM, but may be produced by an alternative higher scale theory.

	Also, in general, the higher order operators we consider do not appear in isolation. For most of the operators that we have considered, they are the only operator added to the model, except where we have combined operators which perform similar functions. The effect of combining operators which directly annihilate is to make the depletion processes more efficient. There may be a roughly two-fold decrease in the size of the interaction required to produce a viable thermal relic. For the Higgs boson interactions, including multiple interactions of varying strength does not change the resultant physics, as there will still in general be a splitting between the neutralinos and chargino. The exception occurs for the higgsinos (winos) where $c_4$($d_5$) is negative relative to $\mu$($M_2$) and in the case of higgsinos is equal to the average of $c_1$ and $c_2$. 

	Here, we have only considered higgsino-like or wino-like dark matter, where the other SUSY neutralino components (if they exist at all) are assumed to be at a significantly higher scale and are integrated out of the equations. Various models exist where the higgsinos and a wino component occur at the same scale, including the so-called ``well-tempered neutralino''~\cite{arkani-hamed06,bharucha17}. While all of the operators we consider also apply to a general four neutralino model, there are no gauge-invariant combinations which mix a wino and a higgsino, though there are Yukawa terms which mix the neutralinos. The additional mixing will impact the particle masses near the low mass viable region. 

	The effective field theory approach allows for simplified models of neutralino dark matter to be analysed without specifying the higher order theory. Contrary to results from the MSSM, where the thermal relic higgsino-like and wino-like dark matter is constrained to be one value, viable thermal relic higgsinos exist in a wide spectrum from a few tens of$\GeV$ to a few tens of$\TeV$. These regions may be constrained in the near future, but composite thermal relic higgsinos remain a key model in the search for dark matter.

%% file: paper.bbl
\providecommand{\href}[2]{#2}\begingroup\raggedright\begin{thebibliography}{10}

\bibitem{zwicky33}
F.~Zwicky, \emph{Die {R}otverschiebung von extragalaktischen {N}ebeln (the
  redshift of extragalactic nebulae)}, {\emph{Heletica Physica Acta} {\bf 6}
  (1933) 110--127}.

\bibitem{roberts75}
M.~S. Roberts and R.~N. Whitehurst, \emph{The rotation curve and geometry of
  {M31} at large galactocentric distances},
  \href{http://dx.doi.org/10.1086/153889}{\emph{The Astrophysical Journal} {\bf
  201} (1975) 327}.

\bibitem{white78}
S.~D.~M. White and M.~J. Rees, \emph{Core condensation in heavy halos: a
  two-stage theory for galaxy formation and clustering},
  \href{http://dx.doi.org/10.1093/mnras/183.3.341}{\emph{Monthly Notices of the
  Royal Astronomical Society} {\bf 183} (1978) 341}.

\bibitem{spergel03}
D.~N. Spergel, L.~Verde, H.~V. Peiris, E.~Komatsu, M.~R. Nolta, C.~L. Bennett
  et~al., \emph{First-year \textit{{W}ilkinson {M}icrowave {A}nisotropy {P}robe
  ({WMAP})} observations: {D}etermination of cosmological parameters},
  \href{http://dx.doi.org/10.1086/377226}{\emph{The Astrophysical Journal
  Supplement Series} {\bf 148} (2003) 175},
  [\href{http://arxiv.org/abs/arXiv:astro-ph/0302209}{{\tt
  arXiv:astro-ph/0302209}}].

\bibitem{clowe04}
D.~Clowe, A.~Gonzalez and M.~Markevitch, \emph{Weak-lensing mass reconstruction
  of the interacting cluster 1{E} 0657-558: {D}irect evidence for the existence
  of dark matter}, \href{http://dx.doi.org/10.1086/381970}{\emph{The
  Astrophysical Journal} {\bf 604} (2004) 596--603},
  [\href{http://arxiv.org/abs/arXiv:astro-ph/0312273}{{\tt
  arXiv:astro-ph/0312273}}].

\bibitem{cirelli06}
M.~Cirelli, N.~Fornengo and A.~Strumia, \emph{Minimal dark matter},
  \href{http://dx.doi.org/10.1016/j.nuclphysb.2006.07.012}{\emph{Nuclear
  Physics B} {\bf 753} (2006) 178},
  [\href{http://arxiv.org/abs/arXiv:hep-ph/0512090}{{\tt
  arXiv:hep-ph/0512090}}].

\bibitem{griest91}
K.~Griest and D.~Seckel, \emph{Three exceptions in the calculation of relic
  abundances}, \href{http://dx.doi.org/10.1103/PhysRevD.43.3191}{\emph{Physical
  Review D} {\bf 43} (1991) 3191}.

\bibitem{mizuta93}
S.~Mizuta and M.~Yamaguchi, \emph{Coannihilation effects and relic abundance of
  higgsino-dominant {LSP}s},
  \href{http://dx.doi.org/10.1016/0370-2693(93)91717-2}{\emph{Physics Letters
  B} {\bf 298} (1993) 120},
  [\href{http://arxiv.org/abs/arXiv:hep-ph/9208251}{{\tt
  arXiv:hep-ph/9208251}}].

\bibitem{drees97}
M.~Drees, M.~M. Nojiri, D.~P. Roy and Y.~Yamada, \emph{Light {H}iggsino dark
  matter}, \href{http://dx.doi.org/10.1103/PhysRevD.56.276}{\emph{Phs. Rev. D}
  {\bf 56} (1997) 276}, [\href{http://arxiv.org/abs/arXiv:hep-ph/9701219}{{\tt
  arXiv:hep-ph/9701219}}].

\bibitem{hisano07}
J.~Hisano, S.~Matsumoto, M.~Nagai, O.~Saito and M.~Senami,
  \emph{Non-perturbative effect on thermal relic abundance of dark amtter},
  \href{http://dx.doi.org/10.1016/j.physletb.2007.01.012}{\emph{Physics Letters
  B} {\bf 646} (2007) 34},
  [\href{http://arxiv.org/abs/arXiv:hep-ph/0610249}{{\tt
  arXiv:hep-ph/0610249}}].

\bibitem{arkani-hamed06}
N.~Arkani-Hamed, A.~Delgado and G.~F. Giudice, \emph{The well-tempered
  neutralino},
  \href{http://dx.doi.org/10.1016/j.nuclphysb.2006.02.010}{\emph{Nuclear
  Physics B} {\bf 741} (2006) 108},
  [\href{http://arxiv.org/abs/arXiv:hep-ph/0601041}{{\tt
  arXiv:hep-ph/0601041}}].

\bibitem{profumo04}
S.~Profumo and C.~E. Yaguna, \emph{Statisical analysis of supersymmetric dark
  matter in the minimal supersymmetric standard model after {WMAP}},
  \href{http://dx.doi.org/10.1103/PhysRevD.70.095004}{\emph{Physical Review D}
  {\bf 70} (2004) 095004},
  [\href{http://arxiv.org/abs/arXiv:hep-ph/0407036}{{\tt
  arXiv:hep-ph/0407036}}].

\bibitem{baer11}
H.~Baer, A.~Lessa, S.~Rajagopalan and W.~Sreethawong, \emph{Mixed
  axion/neutralino cold dark matter in supersymmetric models},
  \href{http://dx.doi.org/10.1088/1475-7516/2011/06/031}{\emph{Journal of
  Cosmology and Astroparticle Physics} {\bf 06} (2011) 031},
  [\href{http://arxiv.org/abs/arXiv:1103.5413}{{\tt arXiv:1103.5413}}].

\bibitem{baer13}
H.~Baer, V.~Barger and D.~Mickelson, \emph{Direct and indirect detection of
  higgsino-like {WIMP}s: {C}oncluding the story of electroweak naturalness},
  \href{http://dx.doi.org/10.1016/j.physletb.2013.08.060}{\emph{Physics Letters
  B} {\bf 726} (2013) 330}, [\href{http://arxiv.org/abs/arXiv:1303.3816}{{\tt
  arXiv:1303.3816}}].

\bibitem{bae15}
K.~J. Bae, H.~Baer and V.~Barger, \emph{Supersymmetry with radiatvely-driven
  naturalness: {I}mplications for {WIMP} and axion searches},
  \href{http://dx.doi.org/10.3390/sym7020788}{\emph{Symmetry} {\bf 7} (2015)
  788}, [\href{http://arxiv.org/abs/arXiv:1503.04137}{{\tt arXiv:1503.04137}}].

\bibitem{bharucha17}
A.~Bharucha, F.~Br\"{u}mmer and R.~Ruffault, \emph{Well-tempered n-plet dark
  matter}, \href{http://dx.doi.org/10.1007/JHEP09(2017)160}{\emph{Journal of
  High Energy Physics} {\bf 2017} (2017) 160},
  [\href{http://arxiv.org/abs/arXiv:1703.00370}{{\tt arXiv:1703.00370}}].

\bibitem{roszkowski18}
L.~Roszkowski, E.~M. Sessolo and S.~Trojanowski, \emph{{WIMP} dark matter
  candidates and searches-current status and future prospects},
  \href{http://dx.doi.org/10.1088/1361-6633/aab913}{\emph{Reports on Progress
  in Physics} {\bf 81} (2018) 066201},
  [\href{http://arxiv.org/abs/1707.06277}{{\tt 1707.06277}}].

\bibitem{kahlhoefer17}
F.~Kahlhoefer, \emph{Review of {LHC} dark matter searches},
  \href{http://dx.doi.org/10.1142/S0217751X1730006X}{\emph{International
  Journal of Modern Physics A} {\bf 32} (2017) 1730006},
  [\href{http://arxiv.org/abs/arXiv:1702.02430}{{\tt arXiv:1702.02430}}].

\bibitem{lux17}
{LUX Collaboration}, \emph{Results from a search for dark matter in the
  complete {LUX} exposure},
  \href{http://dx.doi.org/10.1103/PhysRevLett.118.021303}{\emph{Physical Review
  Letters} {\bf 118} (2017) 021303},
  [\href{http://arxiv.org/abs/arXiv:1608.07648}{{\tt arXiv:1608.07648}}].

\bibitem{xenon17}
{XENON Collaboration}, \emph{First dark matter search results from the
  {XENON1T} experiment},
  \href{http://dx.doi.org/10.1103/PhysRevLett.119.181301}{\emph{Physical Review
  Letters} {\bf 119} (2017) 181301},
  [\href{http://arxiv.org/abs/arXiv:1705.06655}{{\tt arXiv:1705.06655}}].

\bibitem{fermilat15}
{Fermi-LAT Collaboration}, \emph{Searching for dark matter annihilation from
  {M}ilky {W}ay dwarf spheriodal galaxies with six years of {F}ermi {L}arge
  {A}rea {T}elescope data},
  \href{http://dx.doi.org/10.1103/PhysRevLett.115.231301}{\emph{Physical Review
  Letters} {\bf 115} (2015) 231301},
  [\href{http://arxiv.org/abs/arXiv:1503.02641}{{\tt arXiv:1503.02641}}].

\bibitem{ams16}
{AMS Collaboration}, \emph{Antiproton flux, antiproton-to-proton flux ratio,
  and properties of elementary particle fluxes in primary cosmic rays measured
  with the {A}lpha {M}agnetic {S}pectrometer on the {I}nternational {S}pace
  {S}tation},
  \href{http://dx.doi.org/10.1103/PhysRevLett.117.091103}{\emph{Physical Review
  Letters} {\bf 117} (2016) 091103}.

\bibitem{badziak17}
M.~Badziak, M.~Olechowski and P.~Szczerbiak, \emph{Is well-tempered neutralino
  in {MSSM} still alive after 2016 {LUX} results?},
  \href{http://dx.doi.org/10.1016/j.physletb.2017.04.059}{\emph{Physics Letters
  B} {\bf 770} (2017) 226}, [\href{http://arxiv.org/abs/arXiv:1701.05869}{{\tt
  arXiv:1701.05869}}].

\bibitem{fan13}
J.~Fan and M.~Reece, \emph{In wino veritas? {I}ndirect searches shed light on
  neutralino dark matter},
  \href{http://dx.doi.org/10.1007/JHEP10(2013)124}{\emph{Journal of High Energy
  Physics} {\bf 10} (2013) 124},
  [\href{http://arxiv.org/abs/arXiv:1307.4400}{{\tt arXiv:1307.4400}}].

\bibitem{cohen13}
T.~Cohen, M.~Lisanti, A.~Pierce and T.~R. Slatyer, \emph{Wino dark matter under
  seige}, \href{http://dx.doi.org/10.1088/1475-7516/2013/10/061}{\emph{Journal
  of Cosmology and Particle Physics} {\bf 10} (2013) 061},
  [\href{http://arxiv.org/abs/1307.4082}{{\tt 1307.4082}}].

\bibitem{smith01}
D.~Smith and N.~Weiner, \emph{Inelastic dark matter},
  \href{http://dx.doi.org/10.1103/PhysRevD.64.043502}{\emph{Physical Review D}
  {\bf 64} (2001) 043502},
  [\href{http://arxiv.org/abs/arXiv:hep-ph/0101138}{{\tt
  arXiv:hep-ph/0101138}}].

\bibitem{tuckersmith05}
D.~Tucker-Smith and N.~Weiner, \emph{Status of inelastic dark matter},
  \href{http://dx.doi.org/10.1103/PhysRevD.72.063509}{\emph{Physical Review D}
  {\bf 72} (2005) 063509},
  [\href{http://arxiv.org/abs/arXiv:hep-ph/0402065}{{\tt
  arXiv:hep-ph/0402065}}].

\bibitem{bramante16}
J.~Bramante, N.~Desai, P.~Fox, A.~Martin, B.~Ostdiek and T.~Plehn,
  \emph{Towards the final word on neutralino dark matter},
  \href{http://dx.doi.org/10.1103/PhysRevD.93.063525}{\emph{Physical Review D}
  {\bf 93} (2016) 063525}, [\href{http://arxiv.org/abs/arXiv:1510.03460}{{\tt
  arXiv:1510.03460}}].

\bibitem{baer18}
H.~Baer, V.~Barger, D.~Sngupta and X.~Tata, \emph{Is natural higgsino-only dark
  matter excluded?},
  \href{http://dx.doi.org/10.1140/epjc/s10052-018-6306-y}{\emph{The European
  Physical Journal C} {\bf 78} (2018) 838},
  [\href{http://arxiv.org/abs/arXiv:1803.11210}{{\tt arXiv:1803.11210}}].

\bibitem{kowalska18}
K.~Kowalska and E.~M. Sessolo, \emph{The discreet charm of higgsino dark
  matter: {A} pocket review},
  \href{http://dx.doi.org/10.1155/2018/6828560}{\emph{Advances in High Energy
  Physics} {\bf 2018} (2018) 6828560},
  [\href{http://arxiv.org/abs/arXiv:1802.04097}{{\tt arXiv:1802.04097}}].

\bibitem{krall18}
R.~Krall and M.~Reece, \emph{Last electroweak {WIMP} standing: pseudo-{D}irac
  higgsino status and compact stars as future probes},
  \href{http://dx.doi.org/10.1088/1674-1137/42/4/043105}{\emph{Chinese Physics
  C} {\bf 42} (2018) 043105},
  [\href{http://arxiv.org/abs/arXiv:1705.04843}{{\tt arXiv:1705.04843}}].

\bibitem{pierce94a}
D.~Pierce and A.~Papadopoulos, \emph{Radiative corrections to neutralino and
  chargino masses in the minimal supersymmetric model},
  \href{http://dx.doi.org/10.1103/PhysRevD.50.565}{\emph{Physical Review D}
  {\bf 50} (1994) 565}, [\href{http://arxiv.org/abs/arXiv:hep-ph/9312248}{{\tt
  arXiv:hep-ph/9312248}}].

\bibitem{pierce94b}
D.~Pierce and A.~Papadopoulos, \emph{The complete radiative corrections to the
  gaugino and {H}iggsino masses in the minimal supersymmetric model},
  \href{http://dx.doi.org/10.1016/0550-3213(94)00303-3}{\emph{Nuclear Physics
  B} {\bf 430} (1994) 278},
  [\href{http://arxiv.org/abs/arXiv:hep-ph/9403240}{{\tt
  arXiv:hep-ph/9403240}}].

\bibitem{lahanas94}
A.~B. Lahanas, K.~Tamvakis and N.~D. Tracas, \emph{One loop corrections to the
  neutralino sector and radiative electroweak breaking in the {MSSM}},
  \href{http://dx.doi.org/10.1016/0370-2693(94)90211-9}{\emph{Physics Letters
  B} {\bf 324} (1994) 387},
  [\href{http://arxiv.org/abs/arXiv:hep-ph/9312251}{{\tt
  arXiv:hep-ph/9312251}}].

\bibitem{howe12}
K.~Howe and P.~Saraswat, \emph{Excess higgs production in neutralino decays},
  \href{http://dx.doi.org/10.1007/JHEP10(2012)065}{\emph{Journal of High Energy
  Physics} {\bf 10} (2012) 065},
  [\href{http://arxiv.org/abs/arXiv:1208.1542}{{\tt arXiv:1208.1542}}].

\bibitem{gherghetta99}
T.~Gherghetta, G.~F. Giudice and J.~D. Wells, \emph{Phenomenological
  consequences of supersymmetry with anomaly induced masses},
  \href{http://dx.doi.org/10.1016/S0550-3213(99)00429-0}{\emph{Nuclear Physics
  B} {\bf 559} (1999) 27},
  [\href{http://arxiv.org/abs/arXiv:hep-ph/9904378}{{\tt
  arXiv:hep-ph/9904378}}].

\bibitem{feng99}
J.~L. Feng, T.~Moroi, L.~Randall, M.~Strassler and S.~Su, \emph{Discovering
  supersymmetry at the {T}evatron in {$W$}-ino {L}ightest {S}upersymmetric
  {P}article scenarios},
  \href{http://dx.doi.org/10.1103/PhysRevLett.83.1731}{\emph{Physical Review
  Letters} {\bf 83} (1999) 1731},
  [\href{http://arxiv.org/abs/arXiv:hep-ph/9904250}{{\tt
  arXiv:hep-ph/9904250}}].

\bibitem{cheng99}
H.-C. Cheng, B.~A. Dobrescu and K.~T. Matchev, \emph{Generic and chiral
  extensions of the supersymmetric standard model},
  \href{http://dx.doi.org/10.1016/S0550-3213(99)00012-7}{\emph{Nuclear Physics
  B} {\bf 543} (1999) 47},
  [\href{http://arxiv.org/abs/arXiv:hep-ph/9811316}{{\tt
  arXiv:hep-ph/9811316}}].

\bibitem{ibe13}
M.~Ibe, S.~Matsumoto and R.~Sato, \emph{Mass splitting between charged and
  neutral winos at two-loop level},
  \href{http://dx.doi.org/10.1016/j.physletb.2013.03.015}{\emph{Physics Letters
  B} {\bf 721} (2013) 252}, [\href{http://arxiv.org/abs/arXiv:1212.5989}{{\tt
  arXiv:1212.5989}}].

\bibitem{yamada10}
Y.~Yamada, \emph{Electroweak two-loop contribution to the mass splitting within
  a new heavy {$SU(2)_L$} fermion multiplet},
  \href{http://dx.doi.org/10.1016/j.physletb.2009.11.044}{\emph{Physics Letters
  B} {\bf 682} (2010) 435}, [\href{http://arxiv.org/abs/arXiv:0906.5207}{{\tt
  arXiv:0906.5207}}].

\bibitem{ellis00}
J.~Ellis, A.~Ferstl and K.~A. Olive, \emph{Re-evaluation of the elastic
  scattering of supersymmetric dark matter},
  \href{http://dx.doi.org/10.1016/S0370-2693(00)00459-7}{\emph{Physics Letters
  B} {\bf 481} (2000) 304},
  [\href{http://arxiv.org/abs/arXiv:hep-ph/0001005}{{\tt
  arXiv:hep-ph/0001005}}].

\bibitem{hisano05}
J.~Hisano, S.~Matsumoto, M.~M. Nojiri and O.~Saito, \emph{Direct detection of
  the {W}ino and {H}iggsino-like neutralino dark matter at one-loop level},
  \href{http://dx.doi.org/10.1103/PhysRevD.71.015007}{\emph{Physical Review D}
  {\bf 71} (2005) 015007},
  [\href{http://arxiv.org/abs/arXiv:hep-ph/0407168}{{\tt
  arXiv:hep-ph/0407168}}].

\bibitem{nagata15}
N.~Nagata and S.~Shirai, \emph{{H}iggsino dark matter in high-scale
  supersymmetry}, \href{http://dx.doi.org/10.1007/JHEP01(2015)029}{\emph{JHEP}
  {\bf 2015} (2015) 29}, [\href{http://arxiv.org/abs/arXiv:1410.4549}{{\tt
  arXiv:1410.4549}}].

\bibitem{Redi:2020qyf}
M.~Redi, \emph{{Dark Matter from 't Hooft Anomaly Matching}},
  \href{http://arxiv.org/abs/2008.12291}{{\tt 2008.12291}}.

\bibitem{hisano15}
J.~Hisano, D.~Kobayashi, N.~Mori and E.~Senaha, \emph{Effective interaction of
  electroweak-interacting dark matter with {H}iggs boson and its
  phenomenology},
  \href{http://dx.doi.org/10.1016/j.physletb.2015.01.012}{\emph{Physics Letters
  B} {\bf 742} (2015) 80}, [\href{http://arxiv.org/abs/arXiv:1410.3569}{{\tt
  arXiv:1410.3569}}].

\bibitem{nagata15b}
N.~Nagata and S.~Shirai, \emph{Electroweakly interacting {D}irac dark matter},
  \href{http://dx.doi.org/10.1103/PhysRevD.91.055035}{\emph{Phsyical Review D}
  {\bf 91} (2015) 055035}, [\href{http://arxiv.org/abs/arXiv:1411.0752}{{\tt
  arXiv:1411.0752}}].

\bibitem{pospelov00}
M.~Pospelov and T.~{ter Veldhuis}, \emph{Direct and indirect limits on the
  electromagnetic form factors of {WIMP}s},
  \href{http://dx.doi.org/10.1093/mnras/274.3.964}{\emph{Phys. Lett. B} {\bf
  480} (2000) 181}, [\href{http://arxiv.org/abs/arXiv:hep-ph/0003010}{{\tt
  arXiv:hep-ph/0003010}}].

\bibitem{barger11}
V.~Barger, W.-Y. Keung and D.~Marfatia, \emph{Electromagnetic properties of
  dark matter: dipole moments and charge form factor},
  \href{http://dx.doi.org/10.1016/j.physletb.2010.12.008}{\emph{Phys. Lett. B}
  {\bf 696} (2011) 74}, [\href{http://arxiv.org/abs/arXiv:1007.4345}{{\tt
  arXiv:1007.4345}}].

\bibitem{sigurdson04}
K.~Sigurdson, M.~Doran, A.~Kurylov, R.~R. Caldwell and M.~Kamionkowsi,
  \emph{Dark-matter electric and magnetic dipole moments},
  \href{http://dx.doi.org/10.1103/PhysRevD.70.083501}{\emph{Phys. Rev. D} {\bf
  70} (2004) 083501}, [\href{http://arxiv.org/abs/arXiv:astro-ph/0406355}{{\tt
  arXiv:astro-ph/0406355}}].

\bibitem{masso09}
E.~Mass\'{o}, S.~Mohanty and S.~Rao, \emph{Dipolar dark matter},
  \href{http://dx.doi.org/10.1103/PhysRevD.80.036009}{\emph{Phys. Rev. D} {\bf
  80} (2009) 036009}, [\href{http://arxiv.org/abs/arXiv:0906.1979}{{\tt
  arXiv:0906.1979}}].

\bibitem{fitzpatrick10}
A.~L. Fitzpatrick and K.~M. Zurek, \emph{Dark moments and the {DAMA}-{CoGeNT}
  puzzle}, \href{http://dx.doi.org/10.1103/PhysRevD.82.075004}{\emph{Phys. Rev.
  D} {\bf 82} (2010) 075004}, [\href{http://arxiv.org/abs/arXiv:1007.5325}{{\tt
  arXiv:1007.5325}}].

\bibitem{fortin12}
J.-F. Fortin and T.~M.~P. Tait, \emph{Collider constraints on
  dipole-interacting dark matter},
  \href{http://dx.doi.org/10.1103/PhysRevD.85.063506}{\emph{Phys. Rev. D} {\bf
  85} (2012) 063506}, [\href{http://arxiv.org/abs/arXiv:1103.3289}{{\tt
  arXiv:1103.3289}}].

\bibitem{heo10}
J.~H. Heo, \emph{Minimal {D}irac fermionic dark matter with nonzero magnetic
  dipole moment},
  \href{http://dx.doi.org/10.1016/j.physletb.2010.08.035}{\emph{Phys. Lett. B}
  {\bf 693} (2010) 255}, [\href{http://arxiv.org/abs/arXiv:0901.3815}{{\tt
  arXiv:0901.3815}}].

\bibitem{heo11}
J.~H. Heo, \emph{Electric dipole moment of {D}irac fermionic dark matter},
  \href{http://dx.doi.org/10.1016/j.physletb.2011.06.088}{\emph{Phys. Lett. B}
  {\bf 702} (2011) 205}, [\href{http://arxiv.org/abs/arXiv:0902.2643}{{\tt
  arXiv:0902.2643}}].

\bibitem{delnobile12}
E.~{Del Nobile}, C.~Kouvaris, P.~Panci, F.~Sannino and J.~Virkaj\"{a}rvi,
  \emph{Light magnetic dark matter in direct detection searches},
  \href{http://dx.doi.org/10.1088/1475-7516/2012/08/010}{\emph{J. Cosmol.
  Astropart. Phys.} {\bf 2012} (2012) 010},
  [\href{http://arxiv.org/abs/arXiv:1203.6652}{{\tt arXiv:1203.6652}}].

\bibitem{delnobile14}
E.~{Del Nobile}, G.~B. Gelmini, P.~Gondolo and J.-H. Huh, \emph{Direct
  detection of light anapole and magnetic dipole {DM}},
  \href{http://dx.doi.org/10.1088/1475-7516/2014/06/002}{\emph{J. Cosmol.
  Astropart. Phys.} {\bf 2014} (2014) 002},
  [\href{http://arxiv.org/abs/arXiv:1401.4508}{{\tt arXiv:1401.4508}}].

\bibitem{gresham14}
M.~I. Gresham and K.~M. Zurek, \emph{Light dark matter anomalies after {LUX}},
  \href{http://dx.doi.org/10.1103/PhysRevD.89.016017}{\emph{Phys. Rev. D} {\bf
  89} (2014) 016017}, [\href{http://arxiv.org/abs/arXiv:1311.2082}{{\tt
  arXiv:1311.2082}}].

\bibitem{mohanty15}
S.~Mohanty and S.~Rao, \emph{Detecting dipolar dark matter in beam dump
  experiments},  \href{http://arxiv.org/abs/arXiv:1506.06462}{{\tt
  arXiv:1506.06462}}.

\bibitem{geytenbeek17}
B.~Geytenbeek, S.~Rao, P.~Scott, A.~Serenelli, A.~C. Vincent, M.~White et~al.,
  \emph{Effect of electromagnetic dipole dark matter on energy transport in the
  solar interior},
  \href{http://dx.doi.org/10.1088/1475-7516/2017/03/029}{\emph{Journal of
  Cosmology and Astroparticle Physics} {\bf 03} (2017) 029},
  [\href{http://arxiv.org/abs/arXiv:1610.06737}{{\tt arXiv:1610.06737}}].

\bibitem{baer16}
H.~Baer, V.~Barger and H.~Serce, \emph{{SUSY} under seige from direct and
  indirect {WIMP} detection experiments},
  \href{http://dx.doi.org/10.1103/PhysRevD.94.115019}{\emph{Phys. Rev. D} {\bf
  94} (2016) 115019}, [\href{http://arxiv.org/abs/arXiv:1609.06735}{{\tt
  arXiv:1609.06735}}].

\bibitem{ellis98}
J.~Ellis, T.~Falk, G.~Ganis, K.~A. Olive and M.~Schmitt, \emph{Charginos and
  neutralinos in the light of radiative corrections: {S}ealing the fate of
  {H}iggsino dark matter},
  \href{http://dx.doi.org/10.1103/PhysRevD.58.095002}{\emph{Phys. Rev. D} {\bf
  58} (1998) 095002}, [\href{http://arxiv.org/abs/arXiv:hep-ph/9801445}{{\tt
  arXiv:hep-ph/9801445}}].

\bibitem{ho13}
C.~M. Ho and R.~J. Scherrer, \emph{Anapole dark matter},
  \href{http://dx.doi.org/10.1016/j,physletb.2013.04.039}{\emph{Physics Letters
  B} {\bf 722} (2013) 341}, [\href{http://arxiv.org/abs/arXiv:1211.0503}{{\tt
  arXiv:1211.0503}}].

\bibitem{radescu85}
E.~E. Radescu, \emph{On the electromagnetic properties of {M}ajorana fermions},
  \href{http://dx.doi.org/10.1103/PhysRevD.32.1266}{\emph{Physical Review D}
  {\bf 32} (1985) 1266}.

\bibitem{choi07}
S.~Y. Choi, H.~E. Haber, J.~Kalinowski and P.~M. Zerwas, \emph{The neutralino
  sector in the {$U(1)$}-extended supersymmetric {S}tandard {M}odel},
  \href{http://dx.doi.org/10.1016/j.nuclphysb.2007.04.019}{\emph{Nuclear
  Physics B} {\bf 778} (2007) 85},
  [\href{http://arxiv.org/abs/arXiv:hep-ph/0612218}{{\tt
  arXiv:hep-ph/0612218}}].

\bibitem{baer15}
H.~Baer, K.-Y. Choi, J.~E. Kim and L.~Roszkowski, \emph{Dark matter production
  in the early universe: beyond the thermal {WIMP} paradigm},
  \href{http://dx.doi.org/10.1016/j.physrep.2014.10.002}{\emph{Physics Reports}
  {\bf 555} (2015) 1}, [\href{http://arxiv.org/abs/arXiv:1407.0017}{{\tt
  arXiv:1407.0017}}].

\bibitem{belanger02}
G.~B\'{e}langer, F.~Boudjema, A.~Pukhov and A.~Semenov, \emph{{micrOMEGAs}: {A}
  program for calculating the relic denisty in the {MSSM}},
  \href{http://dx.doi.org/10.1016/S0010-4655(02)00596-9}{\emph{Computer Physics
  Communications} {\bf 129} (2002) 103},
  [\href{http://arxiv.org/abs/arXiv:hep-ph/011278}{{\tt arXiv:hep-ph/011278}}].

\bibitem{belanger06}
G.~B\'{e}langer, F.~Boudjema, A.~Pukhov and A.~Semenov, \emph{{micrOMEGAs}:
  {V}ersion 1.3},
  \href{http://dx.doi.org/10.1016/j.cpc.2005.12.005}{\emph{Computer Physics
  Communications} {\bf 174} (2006) 577},
  [\href{http://arxiv.org/abs/arXiv:hep-ph/0405253}{{\tt
  arXiv:hep-ph/0405253}}].

\bibitem{barducci18}
D.~Barducci, G.~B\'{e}langer, J.~Bernon, F.~Boudjema, J.~Da~{S}ilva, S.~Kraml
  et~al., \emph{Collider limits on new physics with micr{OMEGA}s4.3},
  \href{http://dx.doi.org/10.1016/j.cpc.2017.08.028}{\emph{Computer Physics
  Communications} {\bf 222} (2018) 327},
  [\href{http://arxiv.org/abs/arXiv:1606.03834}{{\tt arXiv:1606.03834}}].

\bibitem{belyaev13}
A.~Belyaev, N.~D. Christensen and A.~Pukhov, \emph{{CalcHEP} 3.4 for collider
  physics within and beyond the {S}tandard {M}odel},
  \href{http://dx.doi.org/10.1016/j.cpc.2013.01.014}{\emph{Computer Physics
  Communications} {\bf 184} (2013) 1729},
  [\href{http://arxiv.org/abs/arXiv:1207.6082}{{\tt arXiv:1207.6082}}].

\bibitem{semenov08}
A.~V. Semenov, \emph{{LanHEP}-a package for the automatic generation of
  {F}eynman rules in field theory. {V}ersion 3.0},
  \href{http://dx.doi.org/10.1016/j.cpc.2008.10.012}{\emph{Computer Physics
  Communications} {\bf 180} (2009) 431},
  [\href{http://arxiv.org/abs/arXiv:0805.0555}{{\tt arXiv:0805.0555}}].

\bibitem{komatsu11}
E.~Komatsu, K.~M. Smith, J.~Dunkley, C.~L. Bennett, B.~Gold, G.~Hinshaw et~al.,
  \emph{Seven-year {W}ilkinson {M}icrowave {A}nisotropy {P}robe ({WMAP})
  observations: {C}osmological interpretation},
  \href{http://dx.doi.org/10.1088/0067-0049/192/2/18}{\emph{The Astrophysical
  Journal Supplement Series} {\bf 192} (2011) 18},
  [\href{http://arxiv.org/abs/arXiv:1001.4538}{{\tt arXiv:1001.4538}}].

\bibitem{planck16}
{Planck Collaboration}, \emph{Planck 2015 results. {XIII}. {C}osmological
  parameters},
  \href{http://dx.doi.org/10.1051/0004-6361/201525830}{\emph{Astonomy and
  Astrophysics} {\bf 594} (2016) A13},
  [\href{http://arxiv.org/abs/arXiv:1502.01589}{{\tt arXiv:1502.01589}}].

\bibitem{l389}
{L3 Collaboration}, \emph{A determination of the properties of the neutral
  intermediate vector boson $z^0$},
  \href{http://dx.doi.org/10.1016/0370-2693(89)90703-X}{\emph{Physics Letters
  B} {\bf 231} (1989) 509}.

\bibitem{aleph89}
{ALEPH Collaboration}, \emph{Determination of the number of light neutrino
  species}, \href{http://dx.doi.org/10.1016/0370-2693(89)90704-1}{\emph{Physics
  Letters B} {\bf 231} (1989) 519}.

\bibitem{opal89}
{OPAL Collaboration}, \emph{Measurement of the $z^0$ mass and width with the
  {OPAL} detector at {LEP}},
  \href{http://dx.doi.org/10.1016/0370-2693(89)90705-3}{\emph{Physics Letters
  B} {\bf 231} (1989) 530}.

\bibitem{delphi89}
{DELPHI Collaboration}, \emph{Measurement of the mass and with of the
  $z^0$-particle from multihadronic final states produced in $e^+ e^-$
  annihilations},
  \href{http://dx.doi.org/10.1016/0370-2693(89)90706-5}{\emph{Physics Letters
  B} {\bf 231} (1989) 539}.

\bibitem{baer90}
H.~Baer, M.~Drees and X.~Tata, \emph{Constraints on supersymmetric particles
  from the {CERN} {LEP} data on $z^0$ decay properties},
  \href{http://dx.doi.org/10.1103/PhysRevD.41.3414}{\emph{Physical Review D}
  {\bf 41} (1990) 3414}.

\bibitem{drees88}
M.~Drees, C.~S. Kim and X.~Tata, \emph{Supersymmetry phenomenology and the
  nature of the lightest supersymmetric particle},
  \href{http://dx.doi.org/10.1103/PhysRevD.37.784}{\emph{Phys. Rev. D} {\bf 37}
  (1988) 784}.

\bibitem{ackermann12}
M.~Ackermann, M.~Ajello, W.~B. Atwood, L.~Baldini, G.~Barbiellini, D.~Bastieri
  et~al., \emph{Constraints on the galactic halo dark matter from {F}ermi-{LAT}
  diffuse measurements},
  \href{http://dx.doi.org/10.1088/0004-637X/761/2/91}{\emph{The Astrophysical
  Journal} {\bf 761} (2012) 761},
  [\href{http://arxiv.org/abs/arXiv:1205/6474}{{\tt arXiv:1205/6474}}].

\bibitem{fermilat14}
{Fermi-LAT Collaboration}, \emph{Dark matter constraints from observations of
  25 {M}ilky {W}ay satellite galaxies with the {F}ermi {L}arge {A}rea
  {T}elescope},
  \href{http://dx.doi.org/10.1103/PhysRevD.89.042001}{\emph{Physical Review D}
  {\bf 89} (2014) 042001}, [\href{http://arxiv.org/abs/arXiv:1310.0828}{{\tt
  arXiv:1310.0828}}].

\bibitem{hess16}
{HESS Collaboration}, \emph{Search for dark matter annihilations towards the
  inner galactic halo from 10 years of observations with {H}.{E}.{S}.{S}.},
  \href{http://dx.doi.org/10.1103/PhysRevLett.117.111301}{\emph{Physical Review
  Letters} {\bf 117} (2016) 111301},
  [\href{http://arxiv.org/abs/arXiv:1607.08142}{{\tt arXiv:1607.08142}}].

\bibitem{slatyer10}
T.~R. Slatyer, \emph{The {S}ommerfeld enhancement for dark matter with an
  excited state},
  \href{http://dx.doi.org/10.1088/1475-7516/2010/02/028}{\emph{Journal of
  Cosmology and Astroparticle Physics} {\bf 02} (2010) 028},
  [\href{http://arxiv.org/abs/0910.5713}{{\tt 0910.5713}}].

\bibitem{feng10}
J.~L. Feng, M.~Kaplinghat and H.-B. Yu, \emph{Sommerfeld enhancements for
  thermal relic dark matter},
  \href{http://dx.doi.org/10.1103/PhysRevD.82.083525}{\emph{Physical Review D}
  {\bf 82} (2010) 083525}, [\href{http://arxiv.org/abs/arXiv:1005:4678}{{\tt
  arXiv:1005:4678}}].

\bibitem{cassel10}
S.~Cassel, \emph{Sommerfeld factor for arbitrary partial wave processes},
  \href{http://dx.doi.org/10.1088/0954-3899/37/10/105009}{\emph{Journal of
  Physics G: Nuclear and Particle Physics} {\bf 37} (2010) 105009}.

\bibitem{hisano11}
J.~Hisano, K.~Ishiwata, N.~Nagata and T.~Takesako, \emph{Direct detection of
  electroweak-interacting dark matter},
  \href{http://dx.doi.org/10.1007/JHEP07(2011)005}{\emph{Journal of High Energy
  Physics} {\bf 07} (2011) 005},
  [\href{http://arxiv.org/abs/arXiv:1104.0228}{{\tt arXiv:1104.0228}}].

\bibitem{hill12}
R.~J. Hill and M.~P. Solon, \emph{Universal behavior in the scattering of
  heavy, weakly interacting dark matter on nuclear targets},
  \href{http://dx.doi.org/10.1016/j.physletb.2012.01.013}{\emph{Physics Letters
  B} {\bf 707} (2012) 539}, [\href{http://arxiv.org/abs/arXiv:1111.0016}{{\tt
  arXiv:1111.0016}}].

\bibitem{verkerk92}
P.~Verkerk, G.~Grynberg, B.~Pichard, M.~Spiro, S.~Zylberajch, M.~E. Goldberg
  et~al., \emph{Search for superheavy hydrogen in sea water},
  \href{http://dx.doi.org/10.1103/PhysRevLett.68.1116}{\emph{Physical Review
  Letters} {\bf 98} (1992) 1116}.

\bibitem{mcdonald92}
J.~McDonald, K.~A. Olive and M.~Srednicki, \emph{Relic densities of
  neutralinos},
  \href{http://dx.doi.org/10.1016/0370-2693(92)91431-8}{\emph{Physics Letters
  B} {\bf 293} (1992) 80}.

\end{thebibliography}\endgroup
